\documentclass{aa}
\usepackage{graphicx}
\usepackage{txfonts}

\begin{document}

   \title{Oort Cloud ecology} \subtitle{IV. Exchanging asteroids}

   \author{
        E. Hochart\inst{1} \and
        S. Portegies Zwart\inst{1}
    }
   \institute{
             Leiden Observatory, University of Leiden, 
             Niels Bohrweg 2, 2333 CA Leiden, the Netherlands\\
             \email{hochart@mail.strw.leidenuniv.nl}
    }

   \date{Received 12/02/2026; accepted 21/04/2026}
 
  \abstract
   {}
   {We investigated the influence of cluster environments on asteroids, with particular attention to captured material.}
   {Using numerical methods, we simulated a sub-virial fractally distributed star-forming region and a virialised Plummer-distributed star-forming region. We initialised both models with a virial radius of $0.5$pc and $150$ stars. Stellar populations and their corresponding planetary systems are identical between cluster models. Stars initially hosted $500$ asteroids, and those with mass $M_{*}\leq2.0$ M$_\odot$ were also orbited by one to eight planets. We integrated clusters until $30$ Myr.}
   {The sub-virial fractal cluster exhibits richer dynamics, with asteroids and planets more frequently acquiring high eccentricities and inclinations, along with a larger fraction of captured and rogue objects. Additionally, this cluster configuration has its extreme trans-Neptunian object and Sednoid analogues occupying regions of phase-space in semi-major axis, eccentricity, and inclination commonly frequented by captured asteroids. Although the virialised Plummer model can produce such objects, it is less dynamically active, and the vast majority of asteroids occupying these regions are native rather than captured. Lastly, neither model efficiently forms an Oort Cloud, indicating that Oort Cloud assembly is strongly suppressed in both dynamically hot and more quiescent clusters.}
    {}
   \keywords{Star Cluster -- Earth and Planetary Astrophysics -- Solar and Stellar Astrophysics}

   \maketitle

\section{Introduction}
    At first glance, the Sun appears rather unremarkable. It is a typical G-type star, currently halfway through its main-sequence lifetime, hosting planets on nearly circular and coplanar orbits, parameters that are otherwise susceptible to dynamical interactions \citep[see e.g.][]{2009ApJ...697..458S, 2012MNRAS.419.2448P, 2013MNRAS.433..867H}. Yet, beyond the seemingly pristine planetary orbits, the Solar System minor bodies suggest a more complex dynamical past \citep[see e.g.][]{2004Natur.432..598K, 2024NatAs...8.1380P}. 
    
    Stars typically form in clusters \citep{2003ARA&A..41...57L, 2007MNRAS.380.1271P, 2010ARA&A..48..431P}. As such, the early solar neighbourhood was likely much denser than it is today. Interactions between the Sun and its now long-lost siblings have been proposed to produce some of the peculiar features within the Solar System minor bodies, for instance, the Kuiper cliff at $\sim45$ au \citep{2005Icar..177..246K, 2015MNRAS.453.3157J, 2018ApJ...863...45P} or the existence of highly eccentric and inclined trans-Neptunian objects (TNOs) such as the Sednoids \citep{2004ApJ...617..645B, 2009ApJ...697L..91G, 2014Natur.507..471T, 2014MNRAS.444.2808P, 2019AJ....157..139S, 2025NatAs.tmp..146C}. 
    
    \citet{2015MNRAS.451..144P} introduced the `frozen zone'. Objects within the frozen zone orbit too far from the planets to be dynamically perturbed, but remain near enough to the star to be sheltered from the Galactic tide. As a result, they act as `dynamical fossils', preserving imprints of past encounters or perturbations \citep{2016MNRAS.457.4218J, 2020MNRAS.493.5062V}. They may therefore provide insight into the formation of the Solar System. 
    
    Similarly, the distant Oort Cloud may also preserve traces of the Solar System’s birth environment. \citet{2010Sci...329..187L} suggested that $\gtrsim90\%$ of the Oort Cloud population with semi-major axes between $10^{3}\lesssim a$ [au]$\lesssim2\times10^{5}$ was captured from solar siblings during the cluster phase. 
    
    As the orbits of minor bodies are a potential rich source of information on a star's birth cluster, forthcoming observations from the Legacy Survey of Space and Time \citep[LSST;][]{2019ApJ...873..111I} make for an exciting epoch in Solar System science. Indeed, the LSST is predicted to observe $\sim70\%$ of the main asteroid belt of the Solar System and a more distant asteroidal population within the first two years of the survey \citep{2025AJ....170...99K}. As such, the mission may reveal information on the Solar System's birth cluster. 
    
    Motivated by this, we extend the already rich literature investigating minor-body orbits \citep[e.g.][]{2008CeMDA.102..111R, 2011Icar..214..334F, 2016ApJ...827...52L, 2022MNRAS.512.4078D, 2025A&A...693A.166F} by conducting a numerical investigation of how a star’s birth cluster influences its debris disc, with particular focus on the exchange of planetesimals between planetary systems.
    
\section{Methodology}
    We considered two cluster configurations, modelled on NGC 1333. Each configuration consists of five independent initial realisations for a total of ten runs. We followed and integrated the realisations numerically until $t_{\rm end}=30$ Myr, which roughly corresponds to the expected dissolution time of a $75$ M$_\odot$ cluster \citep{2005A&A...429..173L} and the time at which the Sun left its natal cluster based on the predicted Oort Cloud population \citep{2025A&A...698L..27P}. 
    
    We embedded the clusters in the Galactic potential with Cartesian coordinates $(-8.34,\, 0.0,\, 0.027)$ kpc from the Galactic center with orbital velocity equal to the local circular velocity \citep[$\sim 230$ km s$^{-1}$;][]{2004ApJ...616..872R} plus the Sun's peculiar velocity $(11.4,\, 12.3,\, 7.41)$ km s$^{-1}$ \citep{2010MNRAS.403.1829S} to mimic the Solar orbit.
     
    \subsection{The clusters}
     NGC 1333 is a young ($t_{\rm age}\sim1$ Myr), centrally-concentrated star-forming region composed of $203$ stars \citep{2016ApJ...827...52L}, $150$ of which reside within its inner region ($r\approx1.2$ pc). 

     By comparing simulations with observations, \citet{2017MNRAS.468.4340P} show that the spatial distribution of stars in this inner region is best reproduced if the cluster was initially fractal-like following the prescription of \citet{2004A&A...413..929G}. Their best-fit cluster model has an initial fractal dimension $F_d=1.6$, virial radius $r_{\rm vir}=0.5$ pc, and a virial ratio $Q_{\rm vir}=0.3$, which makes it initially subvirial. We adopted these parameters as the initial conditions for the model `NGC 1333f'\footnote{Associated movie of NGC 1333f available online.} (see table \ref{Tab1:Cluster_IC}). The choice of a subvirial initial state is motivated both by the absence of substructure observed in the region and by evidence that clusters typically form under subvirial conditions \citep[see e.g.][]{2009ApJ...697.1020P, 2010MNRAS.407.1098A}). We sampled stellar masses from a \citet{2001MNRAS.322..231K} initial mass function over the range $m_*\in[0.08,\, 30]\, {\rm M_\odot}$. 
     
     To compare the effect of the initial cluster properties, we considered a second set of runs initialised with a virialised ($Q_{\rm vir}=0.5$), \citet{1911MNRAS..71..460P} distribution (model `NGC 1333p'\footnote{Associated movie of NGC 1333p available online}). These runs used the same particle set as the fractal runs, such that the stellar mass distribution and morphology of planetary systems (planet multiplicity and orbital parameters of both planets and asteroids) are identical. Table \ref{Tab1:Cluster_IC} summarises the cluster initial conditions and characteristic timescales calculated using equations in \citet{1987degc.book.....S} and \citet{2010ARA&A..48..431P}. We calculated the half-mass radius relative to the centre of mass of the stellar population.
     \begin{table}
        \caption{Cluster initial conditions. Values represent mean and $1\sigma$ errors.}
            \centering 
            \begin{tabular}{c c c}
                \hline\hline\\[-0.8em]
                & NGC 1333f & NGC 1333p \\ \\[-0.8em]
                \hline\vspace{-0.75em}\\
               Distribution & Fractal & Plummer \\
               $N_*$ & $150$    & $150$   \\
               $r_{h}$ [pc]  & $0.63\pm0.03$ & $0.40\pm0.03$ \\
               $Q_{\rm vir}$   & $0.3$    & $0.5$    \\ \hline \\[-0.9em]
               $t_{\rm ff}$ [Myr] & $0.62\pm0.02$ Myr & $0.61\pm0.02$ \\
                $t_{\rm cross}$ [Myr] & $4.15\pm0.43$ & $3.07\pm0.15$ \\
                $t_{\rm rh}$ [Myr] & $31.72\pm3.31$ & $23.43\pm1.18$ \\
             \hline
            \end{tabular}
         \tablefoot{Row 1: Spatial distribution. Row 2: Number of stars. Row 3: Half-mass radius. Row 4: Virial ratio. Row 5: Cluster free-fall time. Row 6: Cluster crossing time. Row 7: Cluster half-mass relaxation.}
        \label{Tab1:Cluster_IC} 
     \end{table}
    
    \subsection{The planets}
    The formation of the Oort Cloud requires the presence of wide-orbit giant planets such as the Ice Giants found in the Solar System \citep{1980Icar...42..406F, 1987AJ.....94.1330D} since their relatively large gravitational influence allows the eccentricities of asteroids to reach $e\gtrsim0.999$ within $10$ Myr \citep{2004ASPC..323..371D}. At apoastron, these high-$e$ asteroids experience the effect of Galactic and/or cluster tides \citep{1987AJ.....94.1330D}. These external forces change the asteroid's eccentricity, $e$, in a random walk and can detach the asteroid from the planetary regions. Once detached, the asteroids cease to receive planetary kicks and change their orbits only through the external field. Without this mechanism, the asteroids would be ejected from the system altogether \citep{1986Icar...65...13H}. Over secular timescales of $\mathcal{O}(10^{8}$ yr) \citep{2024MNRAS.527.3054H}, weak tidal torques gradually circularise these orbits, ultimately leading to the formation of the Oort Cloud. This procedure shows that planets play a crucial role in forming an Oort Cloud. 
    
    To account for this, the simulation uses planets synthesised with \texttt{PACE}\footnote{https://github.com/shuohuangGIT/PACE} \citep{2024A&A...689A.338H}. The \texttt{PACE} code adopts the pebble accretion paradigm and accounts for background radiation using the \texttt{FRIED} grid \citep{2018MNRAS.481..452H}. The \texttt{FRIED} grid provides pre-computed mass-loss rates due to external photoevaporation for discs orbiting stars with masses $0.05\leq M_*\, [{\mathrm M}_\odot] \leq1.9$ where the far ultraviolet field strength ranges between $10$-$10^{4}$ G$_0$, and for a variety of disc masses and outer radii. Given the range of stellar masses considered, \texttt{PACE} does not track planet formation around stars with $M_*>2$ M$_\odot$. As such, no planets initially orbit these stars.
    
    The runs initially contained $\sim145$ planetary systems, with planet masses ranging between $3.15\times10^{-5}$ M$_{\rm Jup}$ and $2.33$ M$_{\rm Jup}$. Planets were initially on circular orbits with a $\sigma_i=10^{-3}$ rad inclination dispersion. 
    
    \subsection{The asteroids}
    For NGC 1333f, each star hosts $N_{\rm ast}=500$ asteroids. For NGC 1333p, we increased this to $1000$ per star. Asteroids are represented as test particles in the simulations. With all stars containing a debris disc, the occupation fraction exceeds that from observations \citep[$\gtrsim 80\%$ in young star clusters;][]{2000AJ....120.1396H, 2001ApJ...553L.153H}. This choice is motivated by observations corresponding to a lower bound in the occupation fraction.
    
    The initial asteroid distribution follows a surface density profile $\Sigma\propto r^{-3/2}$  \citep{1981PThPS..70...35H} and a Safronov-Toomre $Q$-parameter $Q=1.0$ \citep{1960AnAp...23..979S, 1972ApJ...178..623T}. The edge of the inner disc corresponds to orbits whose orbital period is equal to one year, a boundary chosen to reduce computational cost. The outer edge of the disc is defined as,
    \begin{equation}
        R_{\rm out}=117\left(\frac{M_*}{\rm M_\odot}\right)^{0.45}\,  {\rm au}, \label{Eqn:Rdisc}
    \end{equation} 
    based on the mass-radius relation of protoplanetary discs found numerically and their observational trends in the young Ophiuchus region \citep{2010ApJ...723.1241A, 2020MNRAS.494.4130H, 2020ApJ...895..126H}.

     \begin{figure}
         \centering\includegraphics[width=.9\columnwidth]{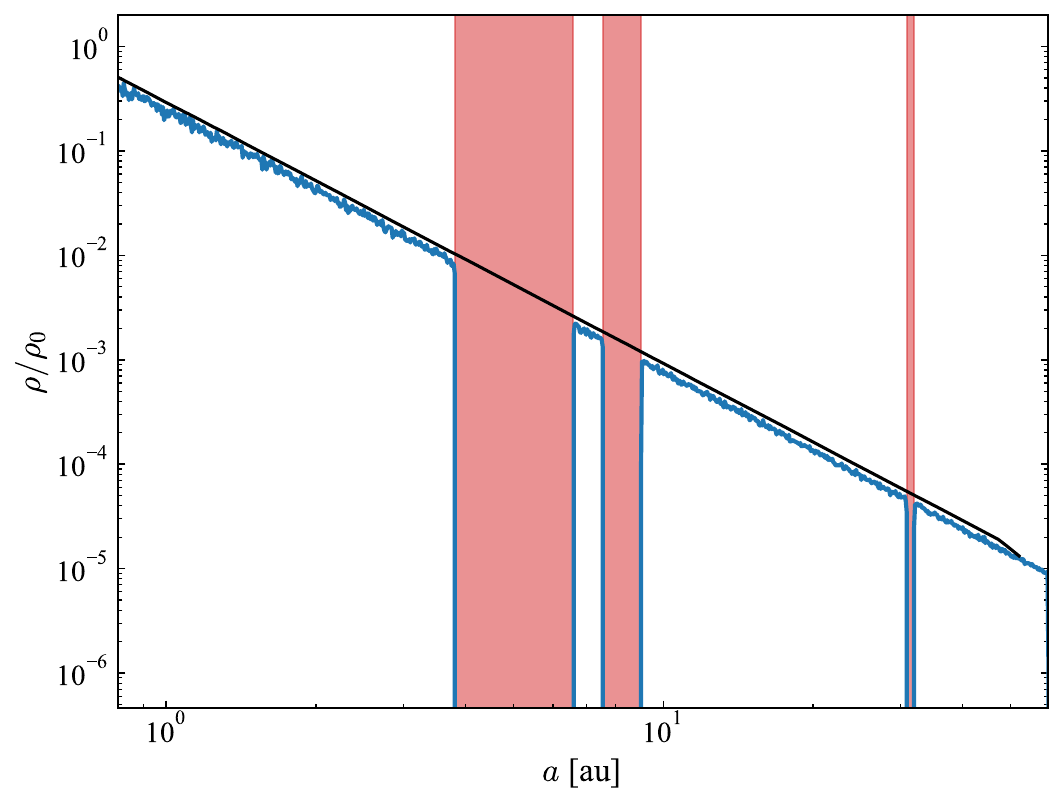}
         \caption{Initial asteroid density profile for a randomly selected system. The black curve shows the theoretical distribution (where $\Sigma\propto r^{-3/2}$). The blue curve shows the asteroid population in that annulus. Shaded red regions indicate locations devoid of asteroids, since we assume planets carve out regions $\pm3$ Hill radii. This system contains three planets.}
         \label{Fig:Astr_Distr}
     \end{figure}
    Planets carve out a gap of $\pm3$ Hill radii along their orbits via dynamical interactions \citep{2023MNRAS.523.4801W}. As a result, any asteroid whose semi-major axis initially lies within this carving region is moved to another part of the disc, and its velocity is appropriately transformed. Figure \ref{Fig:Astr_Distr} shows the initial asteroid distribution for a randomly selected planetary system. In this figure, $N_{\rm ast}$ is increased to reduce statistical noise. As with the planets, we applied a small initial inclination dispersion of $10^{-3}$ radians. After assigning planets and asteroids to a star, we randomised the system's orientation. 
    
    \subsection{The numerics}
        This investigation uses a recently optimised version of \texttt{Nemesis}\footnote{https://github.com/ErwanH29/nemesis} \citep{2019A&A...624A.120V, 2026A&A...709A..30H}, which is a hybrid $N$-body integrator embedded within \texttt{AMUSE} \citep{2013A&A...557A..84P, 2018araa.book.....P}, a library capable in coupling of multi-scale and multi-physics codes. \citet{2026A&A...709A..30H} provide a detailed description of the algorithm. Here, we provide a summary. 
        
    \subsubsection{Challenges}
        Star clusters have crossing times of the order of $1$ Myr, whereas planetary orbits span decades or even days. This wide dynamical range in time presents a major challenge for simulations of planetary systems in star clusters. 
        
        To accurately resolve planetary orbits over long times, it is crucial to suppress secular growth of the energy error. This is typically achieved using symplectic integrators, which consider the system's Hamiltonian and incorporate a shared time-step scheme, ensuring long-term accuracy. However, this same shared time step scheme renders them inefficient for cluster environments since the tightest orbit dictates the time step for all particles, stalling any progress given the $\mathcal{O}(N^2)$ operations required in Newtonian gravity. As a result, codes built for cluster dynamics typically adopt alternative approaches.
        
        Instead, cluster codes generally assign individual time steps to each particle, allowing particles on longer dynamical timescales to have their forces updated less frequently. While this allows efficient modelling of clusters, the approximations make them poorly-suited to resolving small-scale systems such as planetary systems, as evidenced by the drift in energy-error that plagues these algorithms.

        As a result, studies of planetary systems in cluster environments face a trade-off: either employ an optimised cluster code (e.g. a fourth-order Hermite integrator), which efficiently captures global dynamics but poorly resolves planetary motion, or use a symplectic code that accurately models planetary orbits but is computationally prohibitive in environments with a large dynamical range. A promising way to overcome these limitations is through hybrid approaches, which couple specialised integrators under defined assumptions to leverage the strengths of both methods.

    \subsubsection{\texttt{Nemesis}}
        The wide range of length scales present within clusters forms the basis of \texttt{Nemesis}. Since long-range forces act on longer timescales than those nearby, \texttt{Nemesis} assumes that small-scale systems, be they planetary systems, binary stars, triples or sub-clusters, can be decoupled from the global environment and modelled using dedicated symplectic codes. The benefit of this is twofold: the use of a symplectic code for the integration of small-scale systems preserves accurate results, while the removal of small-scale systems in the global system allows the use of $N$-body codes optimised for cluster dynamics. 
        
        In \texttt{Nemesis}, small-scale systems are called `children' and are identified as any conglomerate of particles within
        \begin{equation}
            R_{\rm par} = A\left(\frac{M_{\rm par}}{{\rm M}_\odot}\right)^{1/3}\, {\rm au}. \label{Eqn:Rpar}
        \end{equation}
        Here, $A$ is a scaling factor (set to $100$) and $M_{\rm par}$ is the mass of the corresponding `parent' (defined below). Each child is assigned a dedicated symplectic integrator, which makes the scheme naturally parallelisable. 
        
        To prevent tight orbits from appearing in the cluster integration, children are represented by a parent particle. The parent has phase-space coordinates equal to the centre of mass of the child. Its mass $M_{\rm par}$ is equal to the total mass of the child particles. For example, a child system composed of a binary star system with component masses $M_{1}=M_{2}=1$ M$_\odot$ is represented by a parent particle with mass $M_{\rm par}=2$ M$_\odot$ in the cluster integrator. The parent's radius is set by equation \ref{Eqn:Rpar}. Isolated particles, such as rogue planets or single stars, are treated as self-representing parents. Their mass and phase-space coordinates are taken to be identical to those of the particle itself, and they are directly added to the code that integrates the set of parents. 
        
        The ensemble of parents represents the macroscopic cluster environment, whose dynamical evolution is solved using the fourth-order predictor–corrector Hermite integrator, \texttt{Ph4} \citep{2012ASPC..453..129M}. If two parents are within a distance less than their combined radii, a stopping condition is triggered that interrupts integration. The two parents then merge into a new parent particle, and their associated child particles combine into a new child system. This newly formed child system is subsequently assigned a dedicated integrator to evolve its internal dynamics independently.

        Coupling between the global and local scales (parents and children) is achieved through correction kicks, which are applied to both the set of parents and children at every \texttt{Nemesis} bridge time step, $\delta t_{\rm nem}$, using a second-order Verlet kick-drift-kick method \citep{1967PhRv..159...98V}. For each child, the correction corresponds to the net gravitational force from all parent particles minus that of its own parent representative. Conversely, for parents, the correction is the total force from all children minus that from their associated system. The only gravitational interaction omitted in \texttt{Nemesis} is that between children hosted by different parents, i.e. planets from two well-separated planetary systems \citep[see figure 1 of][]{2019A&A...624A.120V}).
        
        During each \texttt{Nemesis} bridge step, the algorithm checks for the dissolution of any child system using a KD-tree connected-components search with a linking length of $1.2R_{\rm par}$. Particles directly or transitively connected within this distance remain in the same child, so a star and its distant asteroids stay bound, whereas widely separated binaries form distinct parents once beyond the threshold. Subsets containing at least one massive body are represented by a new parent and assigned a dedicated integrator, while the original parent is updated accordingly. Massless clumps (i.e. swarms of asteroids) and isolated massive objects are added directly to the parent code.
        
        \subsubsection{Free parameters}
        In \texttt{Nemesis}, the user is free to choose the integrator. In this study, we evolved the children with the symplectic code \texttt{Huayno} \citep{2012NewA...17..711P}. We used the optimised-Kepler (`\texttt{OK}') scheme of \texttt{Huayno} since it can handle test particles. However, since it cannot resolve collisions, we did not perform an analysis of mergers. As mentioned earlier, parents are evolved with \texttt{Ph4}. The cluster is embedded in a Milky Way-like potential \citep{2015ApJS..216...29B} using the \texttt{BRIDGE} module \citep{2007PASJ...59.1095F} of \texttt{AMUSE}. 
        
        Stellar evolution is handled using the parametric code \texttt{SeBa} \citep{1996A&A...309..179P, 2012ascl.soft01003P} with information on stellar masses and radii relayed to the $N$-body code every $\delta t_{\rm nem}$. The code detects supernova events, at which point the flagged star receives a natal kick in a random direction following \citet{1990ApJ...348..485P}.
        
        The \texttt{Nemesis} bridge time step is fixed to $\delta t_{\rm nem}=10^{3}$ yr. The bridge time step defines the synchronisation interval at which the stellar code communicates with the gravitational integrators, Galactic tidal forces are applied to the cluster, and correction kicks between parent and child subsystems are executed. 
        
\section{Results}
    For the remainder of the paper, we refer objects as `native' (planets or asteroids) if they remain bound at the end of the simulation ($t = 30$ Myr) to the same host to which they were initially bound.
    \subsection{Virialised plummer vs. Sub-virial fractal}
        \subsubsection{Cluster evolution}
            \begin{figure}
                \centering
                \includegraphics[width=.92\columnwidth]{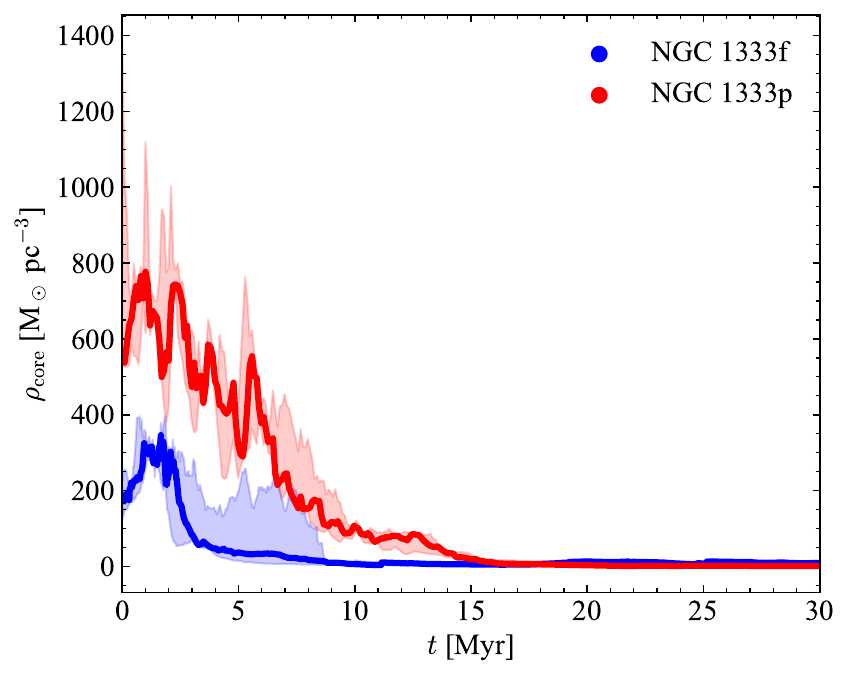}
                \caption{Evolution of cluster density with time. Shaded regions represent the interquartile range.}
                \label{Fig:ClusterDensity}
            \end{figure}
            Models NGC 1333f and NGC 1333p differ greatly in their evolution. Figure \ref{Fig:ClusterDensity} shows how the core density evolves in time between models. We computed the core density following \citet{2003gmbp.book.....H} as
            \begin{equation}
                \rho_{\rm core}=\frac{3(2^{2/3}-1)^{-3/2}}{4\pi}\frac{M_{\rm cluster}}{r_h^3}
            \end{equation}
            where $M_{\rm cluster}$ is the total cluster mass in stars. This formula corresponds to a Plummer model, and as a rough approximation we also use it for the fractal models.

            NGC 1333p has a higher core density, reflecting its initially smooth and centrally concentrated structure. In contrast, NGC 1333f undergoes collapse followed by expansion. Although this collapse temporarily enhances the density, the core density never reaches values attained in NGC 1333p. The maximum core density for NGC 1333f and NGC 1333p are $\rho_{\rm core, max}=405^{+195}_{-44}$ M$_\odot$ pc$^{-3}$ and $\rho_{\rm core, max}=1030^{+578}_{-21}$ M$_\odot$ pc$^{-3}$ respectively.

            Despite this, NGC 1333f is more dynamically active. This apparent discrepancy arises because $\rho_{\rm core}$ is a global quantity that does not capture local density fluctuations. The presence of dense subclumps in NGC 1333f leads to transient local over-densities, increasing the rate of stronger dynamical perturbations compared to the smoother Plummer model. This highlights that global core density alone is insufficient to characterise the level of dynamical activity within a cluster. Instead, local density variations and the degree of substructure play a crucial role in determining the frequency and strength of stellar interactions.

        \subsubsection{Orbital parameters}\label{Sec:OrbParam}
        \begin{figure*}[h]
            \centering                          
            \includegraphics[width=\columnwidth]{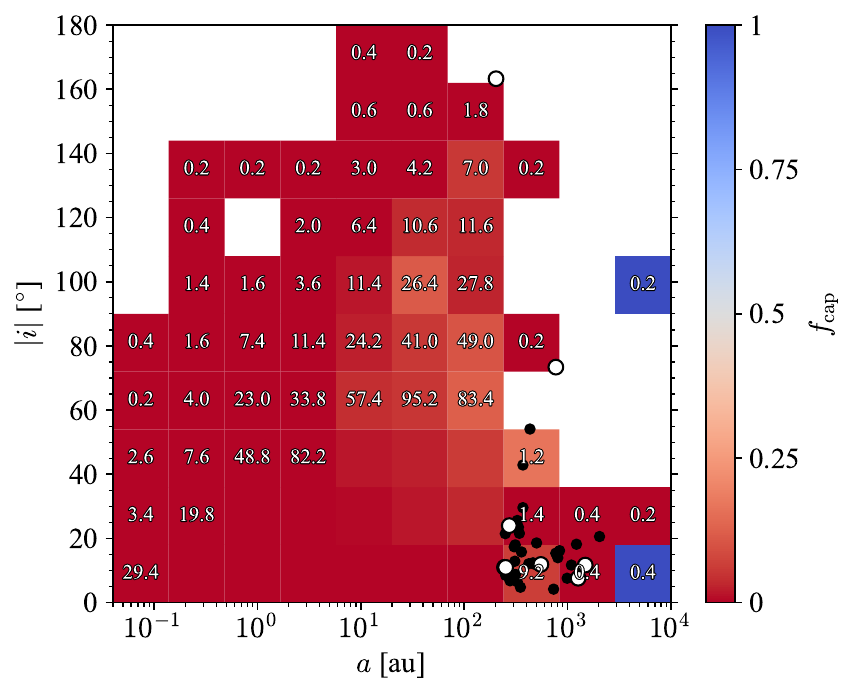}
            \includegraphics[width=\columnwidth]{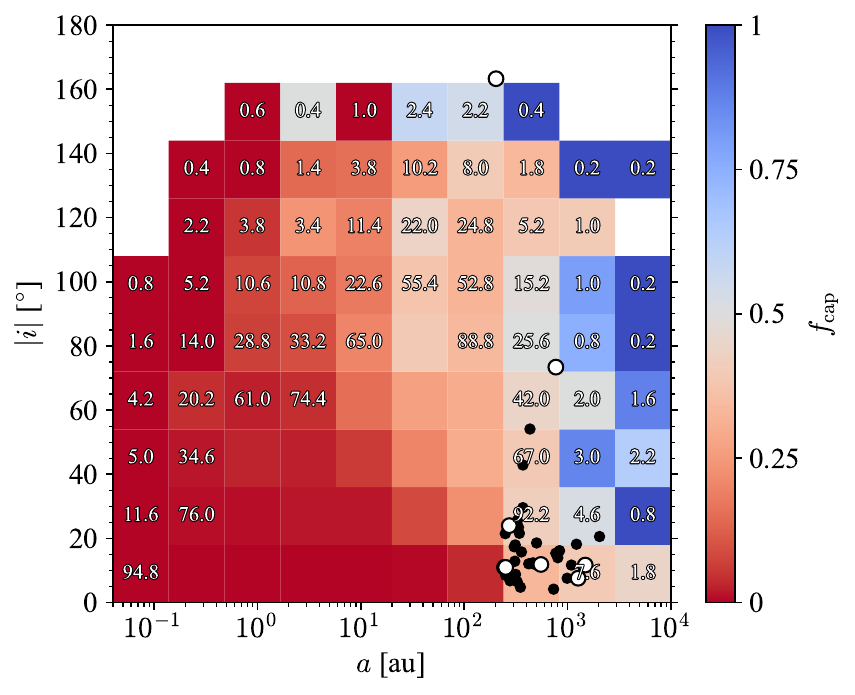}  
            \includegraphics[width=\columnwidth]{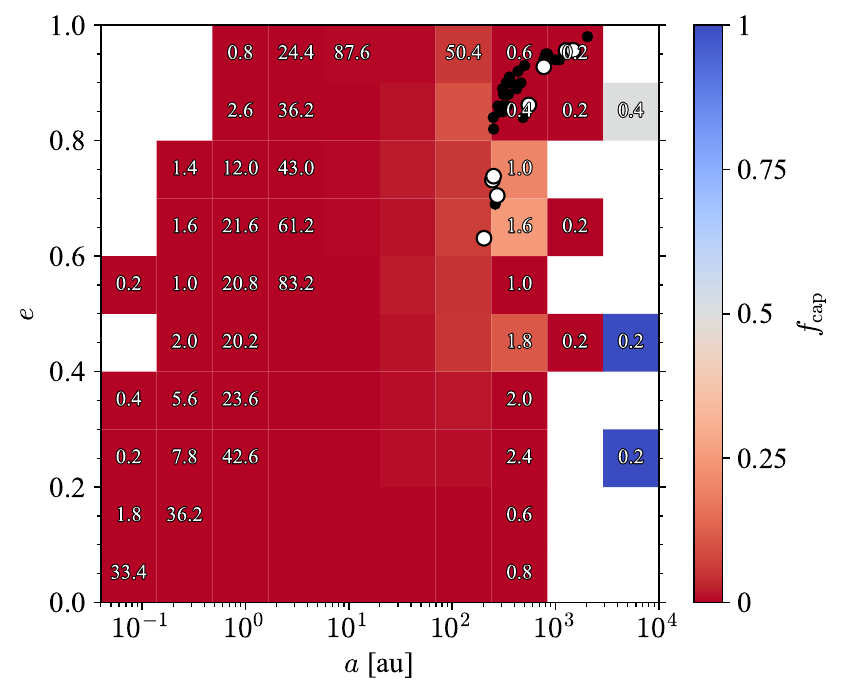}
            \includegraphics[width=\columnwidth]{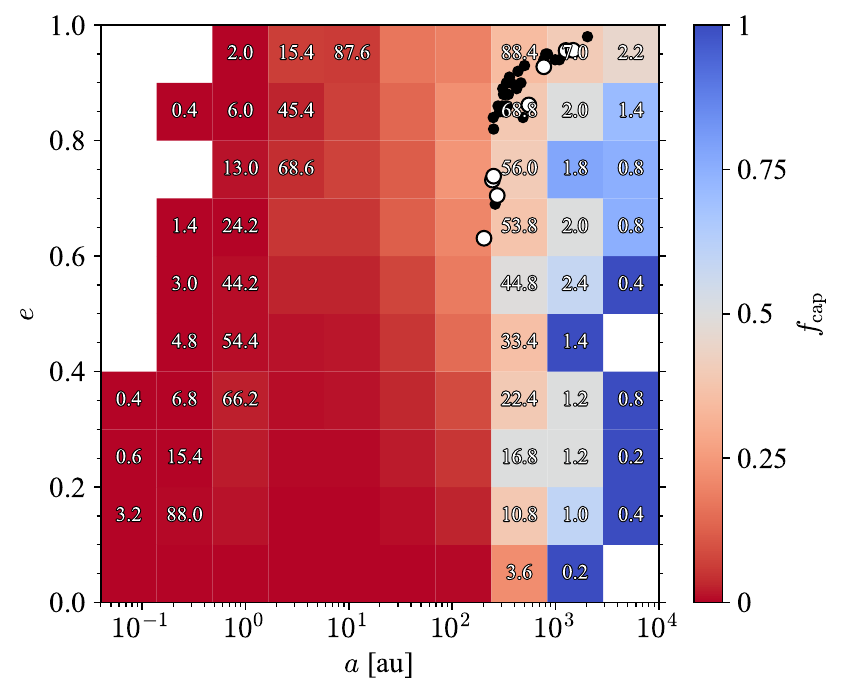}
            \caption{Heatmap of the proportion of captured asteroids in $a$-$i$ (top) and $a$-$e$ (bottom) space after $30$ Myr. Black scatter points indicate the locations of extreme trans-Neptunian objects, while white points indicate Sednoids. Both axes are divided into ten equal-sized bins. Regions with no data are shown in white. Tiles containing fewer than $100$ asteroids per run show their average occupancy. Left column: NGC 1333p. Right column: NGC 1333f.}
            \label{Fig:Heatmap_NGC1333f}
        \end{figure*}
        Before cluster dissolution, a virialised cluster experiences fewer close encounters than a sub-virial one. This difference in behaviour imprints itself on the orbital properties of asteroids in NGC 1333f and NGC 1333p.
        
        Both internal interactions (planet–planet scattering) and external perturbations (stellar flybys) can excite orbital eccentricities, $e$, but changes in inclination, $i$, require non-coplanar encounters. While internal dynamics can alter $i$ \citep{2008ApJ...686..580C, 2008ApJ...686..603J}, the greatest excitations arise from external perturbations. With fewer encounters per star in NGC 1333p, angular-momentum exchange between planetary systems becomes limited. 
        
        Figure \ref{Fig:Heatmap_NGC1333f} shows the distribution of asteroid orbital parameters for both models. Appendix \ref{Sec:AppA} shows the results for each run. The tile colours represent the proportion of asteroids being captured (cf. section \ref{Sec:Captured}). For reference, the Solar System’s extreme trans-Neptunian objects\footnote{https://en.wikipedia.org/wiki/Extreme\_trans-Neptunian\_object\#List} and Sednoids are overlaid on the plot, although caution is warranted when extrapolating to the Solar birth cluster, since this was likely more massive \citep[$N_*\approx10^{3}$–$10^{4}$;][]{2009ApJ...696L..13P, 2010ARA&A..48...47A} than the $N_*=150$ systems modelled here.
        
        The general distribution in $a$ and $i$ of the native asteroids is similar between runs, although NGC 1333f exhibits wider tails in $a$. This is evident from the sharper truncation at $a\approx10^{2}$ au for NGC 1333p. The sharper features in NGC 1333p at $a\approx200$ au and $a\approx4$ au reflect the inner and outer edges of the initialised debris disc. Meanwhile, the extended tails in NGC 1333f result from the more frequent stellar encounters, which diffuse the orbital parameter space occupied by asteroids (and planets). 
        
        \citet{2006Icar..184...59B} argued that a cluster density $\rho\gtrsim 10^{3}$ M$_\odot$ pc$^{-3}$ is required to create Sednoids and that the optimal range is $\rho\approx10^{4}-10^{5}$ M$_\odot$ pc$^{-3}$. While the results of NGC 1333p ($\rho_{\rm core, max}\approx10^{3}$ M$_\odot$ pc$^{-3}$) agree with this statement, with few Sednoids emerging, NGC 1333f with $\rho_{\rm core, max}\approx405$ M$_\odot$ pc$^{-3}$ exhibits a larger Sednoid-like population despite its lower maximum core density. This apparent discrepancy likely stems from differences in planetary architectures (they considered Solar System analogues containing only the giant planets), differences in simulation time ($t_{\rm sim}=3$ Myr versus $30$ Myr here), or their inclusion of gas dynamics, which dampens $e$ and $i$, which makes it harder to scatter objects on wide orbits. Accounting for this, the production of Sednoids is sensitive to initial conditions, and we show how even a low-density cluster can form Sednoids provided it is initially sub-virial and substructured. As such, cluster density alone is not a unique diagnostic.

        Overall, the median asteroid inclination in NGC 1333p is $\langle i\rangle = 0.56^{\circ\,+0.80}_{\,\,\,\,-0.29}$, compared to $\langle i\rangle = 2.99^{\circ\, +6.01}_{\,\,\,-2.04}$ in NGC 1333f. Likewise, for the orbital eccentricities of asteroids, those in NGC 1333p have a median $\langle e\rangle = 0.01_{-0.01}^{+0.03}$ versus $\langle e\rangle = 0.08^{+0.18}_{-0.05}$ for NGC 1333f. The nearly circular orbits in NGC 1333p illustrate how quiescent a virialised Plummer sphere remains compared to its sub-virial, substructured counterpart.
        
        Since this investigation considered a relatively small cluster, gravitational focusing plays a large role, and massive stars have an amplified encounter rate relative to larger clusters \citep{2010A&A...509A..63O}. For reference, FG-type stars experience $\langle N_{\rm enc}\rangle=1360^{+135}_{-251}$ close encounters ($\delta r <10^{4}$ au) over $30$ Myr while M-type stars experience $\langle N_{\rm enc}\rangle=480^{+27}_{-23}$ in NGC 1333f. 
        
        Even so, while we observe dynamical heating of systems orbiting massive stars in NGC 1333f, the same cannot be said for NGC 1333p. The median eccentricity difference between asteroids orbiting OBA-type stars ($M_* > 1.4\,{\rm M_\odot}$) and those around M-type stars ($0.08 \leq M_* < 0.45\,{\rm M_\odot}$) is $\langle \Delta e\rangle = 0.16^{+0.20}_{-0.10}$ in NGC 1333f but only $\langle \Delta e\rangle = 0.01^{+0.00}_{-0.00}$ for NGC 1333p. The small difference between M- and OBA-type stars in NGC 1333p underscores how infrequent encounters are.

        \subsubsection{Kepler dichotomy}
            The Kepler dichotomy is an observational trend identified by the Kepler mission \citep{2010Sci...327..977B}, in which larger $N_{\rm plt}$ systems have lower eccentricity and inclination dispersions \citep{2016PNAS..11311431X, 2018ApJ...860..101Z}. 
            
            Neither model tested replicates this. For NGC 1333p, the dispersion was always too low, while in NGC 1333f, although it reaches the observed values, the dispersion continues to increase when $N_{\rm plt}\geq4$. This discrepancy likely arises because the Kepler mission focused on tight systems (planets with orbital periods $P_{\rm orb}\leq400$ days). Here, planetary systems extend over scales several orders of magnitude larger, making them less susceptible to instabilities and more susceptible to external perturbations. 
            
            Indeed, within compact architectures, even modest dynamical instabilities can trigger planet–planet scattering, making them susceptible to dynamically induced trends. This volatility leads to ejections or mergers, both of which reduce $N_{\rm plt}$ while leaving behind a population with elevated eccentricities and inclinations. We therefore reach the same conclusion as \citet{2015ApJ...807...44P} and \citet{2015ApJ...806L..26V} and posit that there is a survivorship bias within the Kepler sample. High $N_{\rm plt}$ Kepler systems require a nearly pristine environment with minimal internal dynamics to remain stable \citep[see also][]{2025arXiv250106358D}. In contrast, the wider systems modelled here can more easily retain high-$e$, high-$i$ orbits since internal dynamics are less detrimental to the system.

        \subsubsection{Asteroid demographics}
        \begin{figure}
            \centering
            \includegraphics[width=.9\columnwidth]{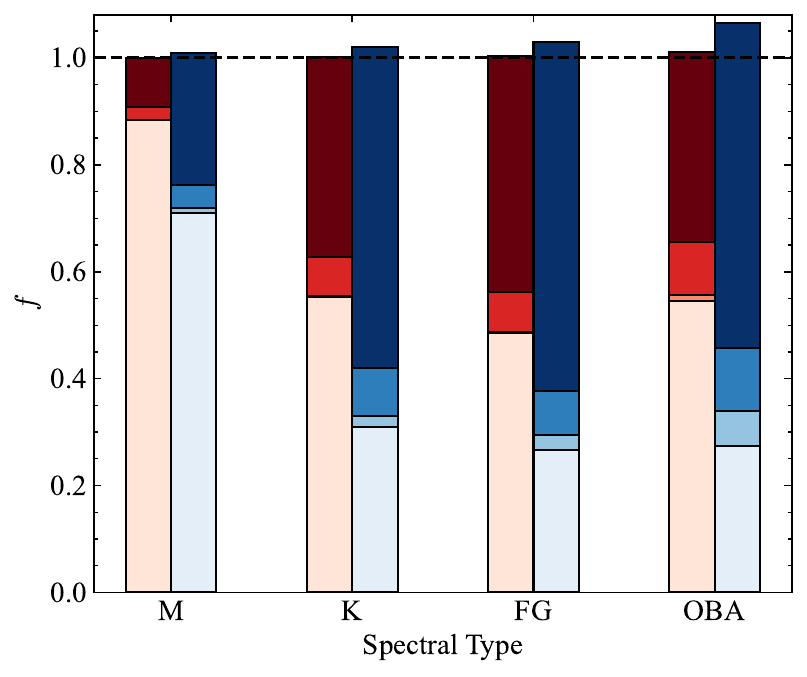}
            \caption{Proportion of asteroid fates binned by stellar spectral types for both models. Red corresponds to NGC 1333p and blue to NGC 1333f. From lighter to darker tones: native fraction (i.e. bound to the same host at the initial and final times), captured fraction, other (removed due to tight orbits), and ejected. Values exceed unity (dashed black line) due to the captured population. Spectral-type bins follow the classification of \citet{1981A&AS...46..193H}, with FG-type and OBA-type stars combined to reduce statistical noise.}
            \label{Fig:AstFate_SpectralTypes_NGC1333f}
        \end{figure}
        
        M-type stars are the most effective at retaining their protoplanetary discs (figure \ref{Fig:AstFate_SpectralTypes_NGC1333f}). This follows naturally from the earlier point that in relatively small clusters, gravitational focusing plays a prominent role in governing stellar encounters and massive stars will more frequently experience close interactions. 
        
        Given the initial mass function used here, M-type stars constitute roughly three-quarters of all cluster members and therefore represent the typical stellar mass. Consequently, K-type stars and beyond have similar survival and ejection rates, while M-type stars, being the least massive and experiencing the fewest encounters, are more efficient at retaining their discs.
        
        Quantitatively, the fraction of surviving native asteroids, defined as $f_{\rm Nat}\equiv N_{{\rm Nat,}f}/N_{\rm Nat,0}$, is $0.78^{+0.03}_{-0.00}$ and $0.68^{+0.07}_{-0.02}$ for M-type stars in NGC 1333p and NGC 1333f respectively, declining to $f_{\rm Nat}=0.53^{+0.02}_{-0.01}$ and $0.23^{+0.10}_{-0.04}$ for OBA-type stars. Here, $N_{\rm Nat,0}$ denotes the initial number of asteroids in the disc and $N_{{\rm Nat,}f}$ denotes the number remaining at the end of the simulation. Within each model, the demographic distribution converges between K-type stars and higher-mass stars ($M_*>0.45\,{\rm M_\odot}$), consistent with \citet{2010A&A...509A..63O}.

        Results from the virialised model (NGC 1333p) closely resemble those of \citet{2006A&A...454..811P}, who simulated an isothermal, virialised Orion-like cluster with $N_*=4000$ over $13$ Myr. Their reported survival fractions, $f_{\rm Nat}\approx0.75$ and $f_{\rm Nat}\approx0.60$ for stars with mass $M_*<1$ M$_\odot$ and $1<M_*$ [M$_\odot$] $<10$, respectively, are comparable to the $f_{\rm Nat}\approx 0.82^{+0.01}_{-0.00}$ and $f_{\rm Nat}\approx 0.59^{+0.07}_{-0.21}$ rates found in NGC 1333p for the same mass ranges (see their figure 2.). This similarity is remarkable given the substantial differences in cluster environment (their clusters having $\rho_{\rm core}\approx2.8\times 10^{3}$ M$_\odot$ pc$^{-3}$), the more compact discs considered here (our discs being $70\%-88\%$ smaller), and our inclusion of planets.
        
        The consistent $f_{\rm Nat}$ values across different virialised Plummer configurations suggest that the survival rate of minor bodies is insensitive to cluster density below a critical threshold that depends on their outer radii. Measurements of interstellar-object number densities or the fraction of dynamically hot asteroids relative to colder ones could provide an indirect test of whether Plummer-type density profiles represent realistic initial conditions for stellar clusters and thus offer new insight into the Sun’s natal environment. 
        
        The more frequent encounters within NGC 1333f also lead to an enhanced exchange of materials between systems, with massive stars being the most efficient asteroid captors. However, stars throughout the mass range are capable of capturing asteroids in the NGC 1333f runs, with $f_{\rm capt}=0.01^{+0.01}_{-0.00}$ and $f_{\rm capt}=0.06^{+0.02}_{-0.01}$ for M- and OBA-type stars, respectively. For NGC 1333p, these rates decrease, with M-type stars not capturing any asteroids and OB-type stars having $f_{\rm capt}=0.01^{+0.00}_{-0.01}$.

        Ejection statistics show a similar trend. The overall ejected asteroid fraction is $f_{\rm ejec}=0.17^{+0.01}_{-0.02}$ in NGC 1333p and $f_{\rm ejec}=0.35^{+0.01}_{-0.02}$ in  NGC 1333f. In both cases, the FG-type stars have the highest rate, with $f_{\rm ejec}=0.49^{+0.07}_{-0.14}$ and $f_{\rm ejec}=0.68^{+0.02}_{-0.07}$ for the two models, respecitvely; the latter value is consistent with \citet{2025A&A...698L..27P}. The high ejection fraction has interesting implications for interstellar interlopers, with a higher number density signifying a more dynamical birth environment. 
        
        Roughly $\sim80\%$ of all ejected asteroids had an initial semi-major axis of $a>30$ au. While chemical measurements of such wanderers within the Solar System may provide information on the outskirts of discs in foreign systems, the non-negligible fraction within $a<30$ au is equally interesting, as these would carry information on planet-forming regions.

    \subsection{Rogue objects}
        \subsubsection{Rogue planets}
        Contrary to \citet{2019A&A...624A.120V}, our results indicate that ejected planets tend to lie in the higher-mass regime. Applying a Kolmogorov-Smirnov test, we find no correlation between the initial planet mass function and the rogue planet mass function ($p=4\times10^{-3}$ and $p=7.6\times10^{-10}$ for NGC 1333f and NGC 1333p, respectively).  
        
        Differences may arise from the initial conditions: they adopt an oligarchic model, in which planets have increasing mass with larger semi-major axes. They also have wide initial orbits, with planetary systems at times reaching $400$ au. Being so wide, massive planets are likely to be ejected. Meanwhile, being more massive and given the tendency for perturbations to propagate inwards within planetary systems \citep{2014prpl.conf..787D, 2017MNRAS.470.4337C, 2018MNRAS.474.5114C}, low-mass inner planets are also likely to be ejected upon crossing orbits. Together, these effects combine to make the rogue planet mass function similar to the complete sampled population.

        Here, a natural outcome of using \texttt{PACE} to synthesise planets is that massive stars tend to host more massive planets, which can be distributed anywhere within the disc (though likely beyond the ice line). Since these massive stars also experience more stellar encounters due to mass segregation and gravitational focusing, their planetary systems are more vulnerable (figure \ref{Fig:AstFate_SpectralTypes_NGC1333f}).
        
        We find no significant correlation between host-star mass and planet multiplicity, nor with system compactness. We quantified the latter using the mutual orbital periods between planets. The absence of correlations between these variables suggests that instabilities within planetary systems are more often seeded by external perturbations. This interpretation is also supported the initial stability of planetary systems, which we verified by evolving each system in isolation for $10$ Myr before conducting runs.
        
        At times, dynamical stellar binaries may form \citep{1975MNRAS.173..729H}. Within a binary system, planets are constantly perturbed, restricting the region of parameter space that characterises stable orbits \citep{1999AJ....117..621H}. Comparing the full stellar mass distribution at the end of the simulation with that of stars residing in binaries in the NGC 1333f runs, we find median masses (with interquartile ranges) of $m_* = 0.23^{+0.19}_{-0.09}$ M$_\odot$ and $m_* = 0.54^{+0.38}_{-0.18}$ M$_\odot$, respectively.

        Observations show a decreasing trend between planet multiplicity and system age. The average planet multiplicity decreases from $\bar{N}_p\approx3$ for systems younger than $1$ Gyr to $\bar{N}_p\approx 1.8$ for those younger than $8$ Gyr old \citep{2023AJ....166..243Y}. A decreasing trend with age can arise from instabilities seeded during the time spent within the birth cluster. Over secular timescales, even small perturbations can drive instability.
        
        For both NGC 1333f and NGC 1333p, asteroids are ejected at similar speeds, with medians $\langle v_{\rm ejec}\rangle=1.10^{+1.60}_{-0.70}$ km s$^{-1}$ and $\langle v_{\rm ejec}\rangle=0.82^{+1.29}_{-0.33}$ km s$^{-1}$ for NGC 1333f and NGC 1333p, respectively, with no statistically significant difference ($p=0.15$). When comparing asteroid and planetary ejection velocities for the same cluster conditions, we obtain $p$-values of $0.008$ and $0.067$ for NGC 1333f and NGC 1333p, respectively. The higher $p$-value in the latter case reflects the stronger influence of internal dynamics in ejecting asteroids, as opposed to the externally driven encounters that dominate in NGC 1333f.

    \subsubsection{Rogue asteroids}
        Across all runs, we identify no free-floating binary planet system or rogue planet-asteroid system. The absence of the latter is noteworthy, as previous numerical studies demonstrate that ejected planets can retain bound companions such as lunar or ring systems \citep{1809.05639, 1712.06500}. These findings have recently been supported by observations indicating a high disc fraction amongst free-floating Jupiter-mass objects \citep{2025A&A...704A.170R}. 

        The lack of such systems in our simulations is likely due to limited resolution and the absence of pre-existing moon or disc structures around planets at initialisation. Indeed, at 0:45 in the video referenced earlier for NGC 1333f, a rogue planet can be seen emerging with what appears to be a remnant disc-like structure.
        
        NGC 1333f has $100^{+5}_{-9}$ (rogue fraction $f_{\rm rogue}=0.257$) rogue planets per run, while NGC 1333p has $46^{+21}_{-4}$ ($f_{\rm rogue}=0.125$). Similar to the argument made for interstellar object rates, observed rates of free-floating planets can help disentangle a cluster's original configuration \citep[see also;][]{2012MNRAS.419.2448P, 2013ApJ...772..142L, 2017MNRAS.470.4337C}. 

    \subsection{Captured objects}\label{Sec:Captured}
        \subsubsection{Captured planets}
            \begin{figure}
                \centering
                 \includegraphics[width=.97\columnwidth]{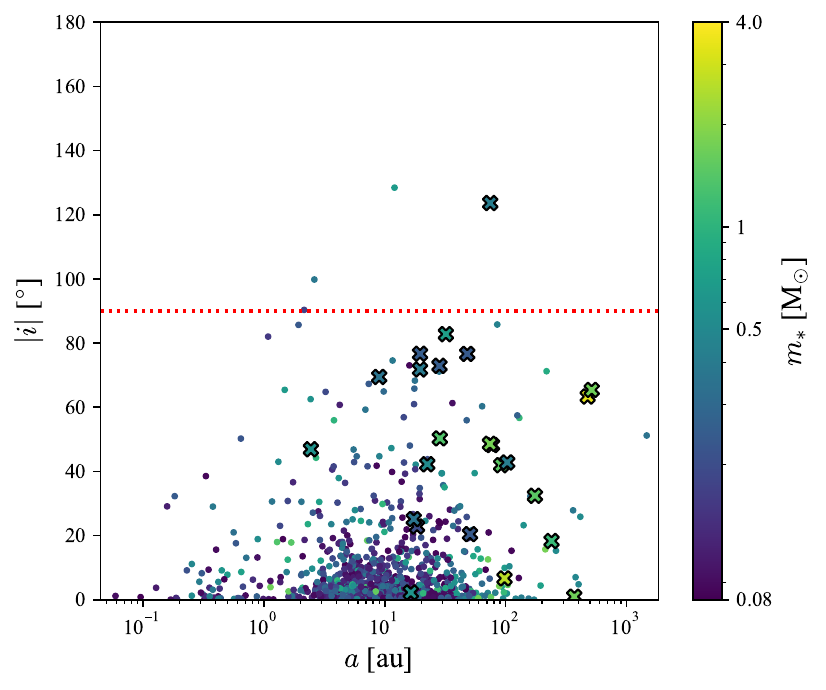}
                \caption{Scatter plot of final semi-major axis versus inclination for planets in all NGC 1333f runs. Dots represent native planets, while crosses represent those that have been captured. Colour represents the mass of the current host star.}
                \label{Fig:CaptPlanets_NGC1333f}
            \end{figure}
            This subsection focuses on NGC 1333f, since no planets are captured in any of the five NGC 1333p runs. For NGC 1333f simulations, the planetary capture rate is $f_{\rm cap}=24/1386=0.02$. Overall, $f_{\rm cap, *}=0.032$ of stars host a captured planet. Given the rogue planet-to-star ratio of $f_{\rm FFP}=2/3$ found here, this agrees with the empirical rate of \citet{2012ApJ...750...83P}, who found $f_{\rm cap, *}=(3-6)\times10^{-2}\times f_{\rm FFP}$. 
            
            Similar to the trend seen for exchanged asteroids, more massive stars are more likely to capture a planet. While the median mass of stars hosting a planet at $t_{\rm sim}=30$ Myr is $\langle m_*\rangle=0.22^{+0.16}_{-0.08}\, {\rm M}_\odot$, the median mass of stars hosting captured planets is $\langle m_*\rangle=0.60^{+0.88}_{-0.24}\, {\mathrm M}_\odot$.
            
            Figure \ref{Fig:CaptPlanets_NGC1333f} shows the final orbital parameters of planets from all NGC 1333f runs. Wide-orbit planets ($a\gtrsim10^{2}$ au) consistently exhibit high eccentricities ($e\gtrsim0.3$), while their inclinations span the full range from coplanar to retrograde. Captured planets (denoted by crosses) occupy similar regions \citep[see also ][]{2022MNRAS.514..920D}, suggesting that measurements of orbital parameters alone are not sufficient to reliably distinguish between a planet scattered outwards and one acquired via capture, although combining parameters can improve confidence.
            
            Captured planets tend to be dynamically hotter. While only a $1\%$ of all planets have inclinations $i>60^{\circ}$, we identify $30\%$ of these high-inclination objects as captured. The median inclinations of native and captured planets are $\langle i \rangle = 2.06^{\circ\ +4.65}_{\,\,\,\,-1.52}$ and $\langle i \rangle = 47.6^{\circ\, +22.5}_{\,\,\,\,-23.0}$, respectively. Similarly, eccentricities are $\langle e \rangle = 0.07^{+0.14}_{-0.05}$ for native planets and $\langle e \rangle = 0.65^{+0.08}_{-0.10}$ for captured planets. 
            
            Planets on very wide orbits ($a > 200$ au) also show similar capture rates with $f_{\rm cap} = 0.31$. When restricted to orbital parameters comparable to the observed planet HD 106906b ($a = 737$ au, $e = 0.44$, $i = 36^\circ$; \citet{2021AJ....161...22N}), planets with $e>0.4$, $a>200$ au, and $i> 30^{\circ}$ have a capture probability $f_{\rm cap} = 0.36$, making HD 106906b a compelling candidate for a non-native origin.
            
            Lastly, across all runs, only one planet lies within the Hill-Oort Cloud (defined as $P_{\rm orb}>10^{4}$ yr, following \citet{2025A&A...698L..27P}), occurring in an NGC 1333f run. For planets with final periastron $r_p>200$ au, we find a wide planet fraction of $f=0.01^{+0.000}_{-0.005}$ in NGC 1333f and $f=0.000^{+0.000}_{-0.000}$ in NGC 1333p.
            
            At the time of writing, a handful of planets have been observed with $a\gtrsim350$ au \citep{2010ApJ...725.1405B, 2014ApJ...780L...4B, 2018AJ....156...57D, 2023MNRAS.519.1688D, 2024AJ....167..253R, 2025A&A...704A.221V}. Of these, six out of ten are circumbinary. Although we find that a third of $a>200$ au planets orbit dynamically formed binaries, these are S-type systems in which the planet orbits only one of the stars. Dynamically forming P-type systems appear difficult, and the observed systems may be primordial in nature. Even so, while large discs have been observed \citep[i.e.][]{2007AJ....134..880H, 2016ApJ...830L..16B, 2020ApJ...898...10T, 2025ApJ...990L...8V, 2026ApJ...996...45M}, discs $>200$ au remain uncommon in young clusters \citep{2005A&A...441..195V, 2017ApJ...851...85B}.

        \subsubsection{Captured asteroids}
\begin{figure}
                \centering
                \includegraphics[width=.92\columnwidth]{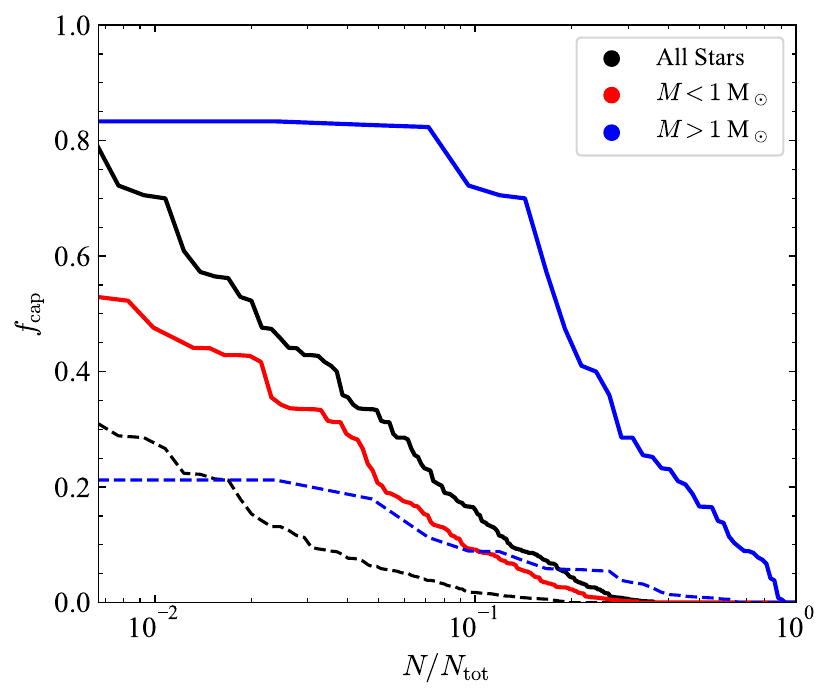}
                \caption{Fraction of stars hosting $f_{\rm cap}$ captured objects for NGC 1333f. The data only includes stars with at least $N_{\rm ast}>10$ at the end time. Solid lines include all asteroids bound to a star. Dashed lines include only asteroids with periastron $r_p<10$ au. The relevant curve for $M<1$ M$_\odot$ is nearly identical to the global curve and omitted to reduce clutter.}
                \label{Fig:ExoticPop_NGC1333f}
            \end{figure}

            Figure \ref{Fig:ExoticPop_NGC1333f} shows the cumulative distribution of contamination levels, analogous to an h-index since the $x$ axis sets a minimum captured fraction, while the $y$ axis gives the fraction of systems that exceed that level. Only systems with $N_{\rm ast}>10$ are considered to suppress small-number noise (Appendix C).
            \begin{figure}
                \centering
                \includegraphics[width=.92\columnwidth]{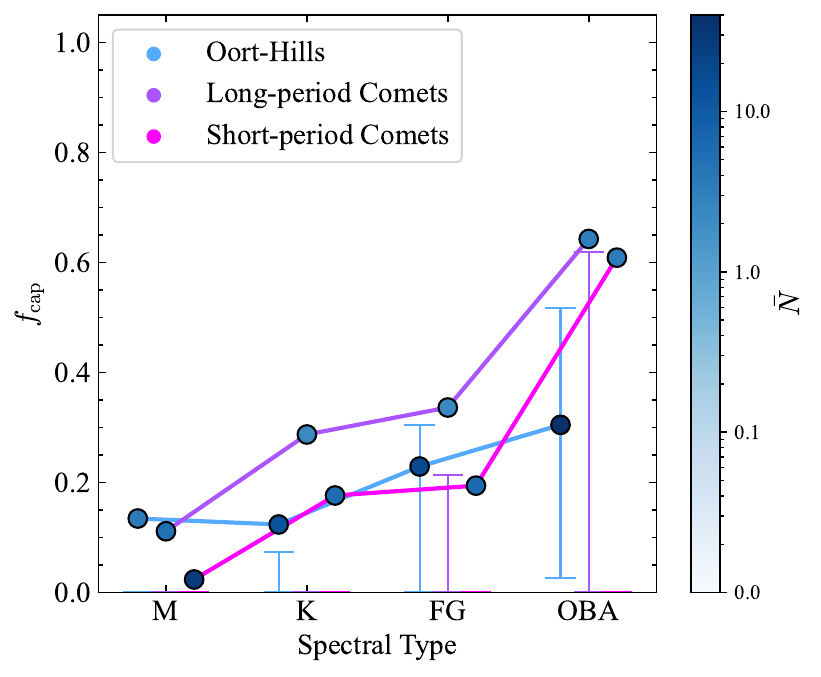}
                \caption{Fraction of captured population for various asteroid families. Hills-Oort objects have $P_{\rm orb} \lesssim5.2$ Myr and $r_p\geq60$. Long-period comets have $e>0.2$, $r_p<10(M_*/$ M$_\odot)^{1/3}$ au and $P_{\rm orb}>200$ yr. Short-period comets have the same $e$ and $r_p$ criteria but have $P_{\rm orb}<200$ yr. Scatter-point colours indicate the mean number of asteroids per spectral classification bin. Bars represent the $25\%$ and $75\%$ quartile range. Cases where the mean exceeds the upper quartile range indicate that outlier stars dominate the mean. Colours also indicate the mean population of each family per stellar system.}
                \label{Fig:AstTypes_NGC1333f}
            \end{figure}
            
            Focusing on the global results (black curve), $13\%$ of the stellar population host systems with at least $>10\%$ of captured material and $2.2\%$ have a $>50\%$ captured population. If we include systems with $N_{\rm ast}<10$, these values increase to $18\%$ and $6\%$, respectively. Restricting back to systems with $N_{\rm ast}>10$, NGC 1333p exhibits reduced capture efficiency: $1.4\%$ of its stars contain at least $10\%$ captured material, and $0.5\%$ exceed $50\%$ rate. 
            
            When considering $M>1$ M$_\odot$ stars, $65\%$ and $21\%$ contain $>10\%$ and $>50\%$ captured material, respectively, highlighting the efficiency of cluster environments in transferring material between systems. The dashed curves in the figure consider asteroids with periastron $r_p<10$ au. For $M_*>1$ M$_\odot$, $\sim1\%$ of stars have at least $10\%$ of their asteroidal population that reaches the inner region that is non-native.
            
            Figure \ref{Fig:AstTypes_NGC1333f} breaks the captured population down by dynamical family for NGC 1333f. Outliers occasionally skew the values, but two trends emerge: capture fractions increase with stellar mass, and a modest fraction of long-period comets around FG stars are non-native ($f\sim0.0$–$0.2$). Over time, we expect this fraction to decline as internal scattering and Galactic tides reshape the outer reservoir \citep{1979MNRAS.187..445Y}.
    
            The Oort-Hill cloud (OHC) is defined in a Solar System-centric way, since asteroids require periastrons $r_p\geq60$ au, a value dependent on Neptune's orbit. The progressively darker tones indicate a more massive OHC around higher-mass stars, consistent with \citet{2012Icar..217....1B}. 

            We note that previous studies propose that the OHC is built primarily by external perturbations (cluster tides and stellar encounters) since these act to raise the perihelia of an asteroid and decouple them from the system \citep{2008Icar..196..274B, 2015MNRAS.451..144P, 2019MNRAS.490...21H, 2025A&A...698L..27P}. Given this mechanism, denser clusters are expected to produce such orbits more easily \citep{2012Icar..217....1B}. Similar to our earlier results regarding Sednoids (section \ref{Sec:OrbParam}), we find that the OHC population is not purely dependent on cluster density, but also on sub-viriality and/or the initial stellar distribution.
            
            The importance of a cluster environment is also apparent with the contrasting population of the OHC and the Öpik-Oort Cloud (ÖOC). While the ÖOC population is continuously replenished through planetary interactions \citep{2004ASPC..323..371D} and the Galactic tide \citep{1986Icar...65...13H}, the OHC lies too deep within its host star's potential to be influenced by the Galactic field and thus remains largely unaffected. Over time, the relative proportion of ÖOC-to-OHC population will decrease. Solar System calculations approximate the OHC to be between two \citep{2025A&A...698L..27P} and five \citep{1987AJ.....94.1330D} times more massive than the ÖOC.
            
            The non-zero rate of captured material crossing the inner regions of planetary systems and the potential presence of captured short- and long-period comets (figure \ref{Fig:AstTypes_NGC1333f}) provides an interesting avenue for lithopanspermia, although this would require the ingredients for life to develop on Myr timescales. Equally interesting is the possibility of deriving information on the composition of discs in other systems. If the prominence of asteroid capture holds in larger clusters, then comets originating from other stars should exist in a majority of extrasolar systems, including our own. 
            
            Considering our low resolutions \citep[$N_{\rm ast}=500$ per star versus a predicted Solar System Oort Cloud population of $\sim5\times10^{11}$, ][]{2005ApJ...635.1348F, 2010Sci...329..187L, 2024come.book...97K}, even with a low capture rate, a handful of such objects will be present in our Solar System. These objects could serve as astronomical Rosetta stones that record the chemical environment of the Sun's stellar siblings and provide insights into planet formation in extrasolar systems. Whether compositional differences exist between debris discs born from the same molecular cloud remains an open question, but observations reveal substantial spatial variations in ice abundances within a single cloud \citep{2006A&A...453L..47P, 2017MNRAS.467.4753N, 2025NatAs...9..883S}.
            
\begin{figure}
                \centering
                \includegraphics[width=.85\columnwidth]{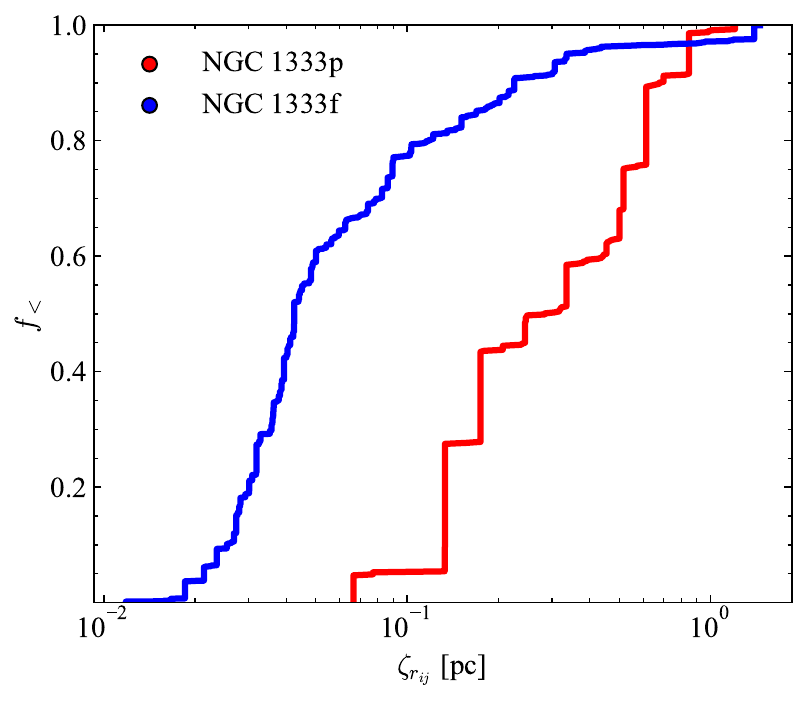}
                \caption{Two-point correlation function of initial distance between an asteroid's original host star and its eventual captured star.}
                \label{Fig:TwoPoint}
            \end{figure}

           However, exchanged asteroids are typically sourced from a star’s neighbours. For NGC 1333f, half of all captured asteroids originate in systems whose initial separation from their final host does not exceed $0.04$ pc (Figure \ref{Fig:TwoPoint}). The jagged features indicate that captures occur in episodic batches rather than through the capture of passing interstellar objects. As the cluster dissolves, early-time encounter-driven exchange decreases, and late-time captures may instead involve bodies captured by chance trajectory alignment (figure 3a of \citet{2010Sci...329..187L}).
            
            Figure \ref{Fig:Heatmap_NGC1333f} shows the fraction fraction of captured asteroids occupying particular $(a,|i|)$ and $(a,e)$ values. As shown by the numbers on each tile, NGC 1333p produces far fewer wide orbits ($a\gtrsim250$ au) and retrograde orbits, with only a handful occupying these regions. In contrast, NGC 1333f efficiently populates wide orbits, with many $a\gtrsim100$ au asteroids originating from other stars, consistent with the idea that Sednoids form a captured population \citep{2004Natur.432..598K, 2015MNRAS.453.3157J}. In NGC 1333p, wide-$a$ asteroids are almost always scattered from the inner system.
    
            While the regions in ($a,i$)- and ($a,e$)-space occupied by captured and native asteroids overlap, their statistics differ substantially. Captured asteroids typically occupy higher inclinations and eccentricities, as well as wider semi-major axes than the native population. 
            
            For FG-type stars in the NGC 1333f runs, the median eccentricies of the captured and native population are $\langle e\rangle=0.56^{+0.21}_{-0.19}$ and $\langle e\rangle=0.23^{+0.23}_{-0.15}$, respectively. Both cases have values similar to those found in \citet{2019MNRAS.490...21H}, although the contrast between the occupied regions of native and captured asteroids is more apparent here (see their figure 5). This difference likely arises because the inner edge of the primordial debris disc considered here being closer the host star. Consequently, a larger proportion of the native asteroid population is sheltered from the external environment. 
            
            Systems that do not experience close encounters (i.e. with nearest stellar encounter distances $\delta r_{\min}>10^{3}$ au) may also host high-$e$, high-$i$ asteroids. This results from internal dynamics, namely the presence of planets whose gravitational influence helps increase $e$ and $i$ \citep[see also][]{2023MNRAS.523.4801W}. While planets can allow for low-angular-momentum orbits, they can also reduce the population of asteroids whose eccentricity approaches unity, since sufficiently massive planets and/or those occupying wide orbits act as additional channels that eject eccentric asteroids from the system.

    \section{Inhibiting Oort Cloud formation in cluster environments}
        Consistent with previous literature \citep[i.e.][]{2015AJ....150...26H, 2021A&A...652A.144P, 2025A&A...698L..27P, 2025ApJ...987...29O}, we find that Oort Cloud formation is suppressed in a star-cluster environment. In what follows, we classify an object as an Oort Cloud member if its semi-major axis lies in the range $1100-5\times10^{4}$ au, following the definition used in \citet{2010Sci...329..187L}. 
        
        Overall, the formation rates found here agree within a factor of two with those of \citet{1990CeMDA..49..265Z}, who considered a virialised $N_*=100$ star cluster, but remain below what would be required if the Oort Cloud had a predominantly extrasolar origin, as argued by \citet{2010Sci...329..187L}. \citet{2010Sci...329..187L} numerically simulated an initially virialised cluster with $N_*=30$ -- $300$ members that became super-virial after gas expulsion at $3$ Myr. To illustrate the suppression of Oort Clouds while a Solar-like star remains embedded in its natal cluster, we performed a first-order estimate of the required protoplanetary disc mass.

        Focusing on NGC 1333f since this gives the greatest Oort Cloud production efficiency, FG-stars have $N_{\rm OC,\, sim}\approx0.29$ Oort Cloud objects at the end of the simulation. The production efficiency scales with stellar mass, with a typical M-star ending with $N_{\rm OC,\, sim}\approx 0.04$ Oort Cloud objects.
        
        The present-day Oort Cloud is theoretically predicted to contain $N_{\rm OC, \, obs}\sim5\times10^{11}$ objects larger than $1$km in size. If the Oort Cloud was primarily assembled while the Sun was still embedded in its natal cluster \citep[a requirement for a capture-dominated origin; see ][]{2022AsBio..22.1429A}, and assuming that the formation efficiency measured in our simulations applies to the Solar System, then the size of the primordial debris disc, $N_{\rm ast,\, obs}$ can be estimated by equating formation efficiencies. That is,
        \begin{equation}
            \left(\frac{N_{\rm OC}}{N_{\rm ast}}\right)_{\rm sim}=\left(\frac{N_{\rm OC}}{N_{\rm ast}}\right)_{\rm obs}.
        \end{equation}
        Re-arranging for $N_{\rm ast, \, obs}$ and substituting the relevant values ($N_{\rm ast,\, sim}=500$, $N_{\rm OC,\, sim}=0.29$ and $N_{\rm OC,\, obs}=5\times10^{11}$), we obtain an estimated initial debris disc population of 
        \begin{equation}
            N_{\rm ast,\,obs} =N_{\rm ast,\, sim}\frac{N_{\rm OC,\, obs}}{N_{\rm OC, \, sim}}\approx8.6\times10^{14}.
        \end{equation}
        Assuming an average asteroid mass of $\langle m_{\rm ast}\rangle=4\times10^{13}$ kg \citep{1996ASPC..107..265W}, this corresponds to an initial debris disc mass of $M_{\rm disc}\approx0.02 M_\odot$.
        
        This estimate represents a lower limit as it assumes a frozen Oort Cloud population after cluster dispersal. In reality, the Oort Cloud population continues to erode due to Galactic tides \citep{1985AJ.....90.1548H, 2024A&A...683A.146I, 2025A&A...699A.248I}. Indeed, for a Sun-like star on a Solar-like orbit, the Oort Cloud population has a half-life of $\sim4$ Gyr \citep{2018MNRAS.473.5432H, 2025A&A...698L..27P}. Accounting for this effect, $M_{\rm disc} \approx0.04$ M$_\odot$. 
        
        Observationally, protoplanetary discs around Solar-mass stars typically have mass ratios in the range $M_{\rm disc}/M_*\approx(0.07-5)\times10^{-3}$ \citep{2013ApJ...771..129A, 2016ApJ...831..125P}. As such, the assumption that the Oort Cloud formed in the cluster and is predominantly sourced by non-native asteroids would require disc masses at least eight times larger than those typically observed. When considering the sensitivity of Oort Cloud formation to cluster morphology \citep[see e.g.][]{2006Icar..184...59B, 2008Icar..196..274B}, this factor increases. 
        
        Encounters occur more frequently in denser clusters. While this increases the probability of asteroid capture, which predominantly occupy wide orbits, these same encounters also strip the existing Oort Cloud population, with the net effect being to hinder Oort Cloud formation \citep[see also][]{2018MNRAS.473.5432H, 2021PSJ.....2...53N, 2022AsBio..22.1429A}. Moreover, the stronger tidal field of the cluster further hinders Oort Cloud formation because it more efficiently erodes the outer reservoir. This tendency is illustrated in figures 10 and 11 of \citet{2006Icar..184...59B}. Consequently, the already stringent disc-mass requirement derived above becomes even more difficult to satisfy in dense birth environments, such as those expected for the Sun.
        
        The high capture fraction found in \citet{2010Sci...329..187L} likely stems from several assumptions that favour it:
        \begin{enumerate}
            \item Asteroids start on wide ($10^{3}<a\,[\mathrm{au}]<5\times10^{3}$), weakly bound orbits, maximising the capture cross-section \citep{1975MNRAS.173..729H, 2020MNRAS.493L..59H, 2021PSJ.....2..217N, 2022MNRAS.514..920D}.
            \item Wide orbits also make stripping via random walks in energy more efficient. This generates a low velocity dispersion between stripped asteroids and stars. Combined with instantaneous cluster expansion, many asteroids follow trajectories similar to the dispersing stars and become re-captured on wide orbits \citep{2010MNRAS.404.1835K, 2011MNRAS.415.1179M}.
            \item Planetary systems are ignored in these calculations, eliminating a perturbing source that might otherwise unbind captured orbits.
        \end{enumerate}
        To emphasise point two, \citet{2010Sci...329..187L} found that $18\%$ of all Oort Cloud asteroids were bound at $3$ Myr; this increased to $71\%$ at $3.5$ Myr.

    \section{Limitations}
        The symplectic integrator employed here does not resolve physical collisions or tidal forces. Neglecting collisions artificially enhances the ejection rate and produces a dynamically hotter surviving population, enhancing the production of high-$a$ objects. This bias is partially mitigated, as a fraction of asteroids that should merge instead get ejected and thus are not considered in the bound population. 

        Similarly, \citet{2001Natur.409..589S} showed that accounting for asteroid-asteroid collisions reduces the number of detached objects by dampening $e$ and $i$. Inclusion of this effect would strongly affect cometary populations. Disc–planetesimal interactions would similarly damp $e$ and $i$ and drive planetary migration \citep{1980ApJ...241..425G, 2010apf..book.....A}. These mechanisms would yield dynamically colder architectures.
        
        Inclusion of tidal effects would both increase the number of collisions and enhance the number of survivors. The latter occurs because high-$e$ asteroids near ejection can circularise at periastron, becoming stably bound close to the host star. This process is analogous to how external tidal fields detach high-$e$ orbits to form an Oort Cloud. Since we focus on the exchanged population, which predominantly occupies high-$a$ orbits (figure \ref{Fig:Heatmap_NGC1333f}), the omission of collisions and tides is unlikely to change our principal conclusions.

        Another simplifying assumption is the omission of gas. Although gas drag has little influence at wide orbits \citep{2003ApJ...592..986P}, it dampens eccentricities and inclinations within the inner disc \citep{2006Icar..184...59B}. As a result, the dynamically hot inner-region planetesimals should have a reduced population, once more impacting the cometary population the most. A massive circumstellar disc can further dampen scattering at wider separations \citep{2010ApJ...711..772R, 2023MNRAS.526.1987F}, lowering the production of detached objects while slightly enhancing capture.
        
        \citet{2022AsBio..22.1429A} find that including gas can enhance capture rates by $\sim10\%$. Although this effect is minor for clusters already in equilibrium, where close encounters are less frequent, it could boost the capture rate in clusters born with substructure. Indeed, assuming that substructures are erased within a crossing time \citep{1972A&A....21..255A, 1997MNRAS.284..785G, 2004A&A...413..929G}, most stellar encounters occur before gas dissipation, which itself acts over $\sim0.1-1$ Myr timescales \citep{1995ApJ...450..824S, 1996AJ....111.2066W, 2006A&A...453L..47P, 2007ApJ...662.1067H, 2024A&A...689A.338H}. 

        Finally, the low-$N$ nature of the simulations introduces statistical noise and run-to-run variability, particularly in the sub-virial NGC 1333f ensemble. This is mitigated by averaging over five realisations per model, yet still falls drastically short of realistic values.

\section{Conclusions}
    We numerically investigated the exchange of protoplanetary disc material within star clusters using \texttt{Nemesis}. We modelled two star clusters, with each considering five sets of initial conditions for a total of ten runs. Clusters contain $150$ stars and approximately $145$ planetary systems. One model adopts a virialised Plummer configuration (NGC 1333p), representative of slightly evolved systems, while the other assumes a sub-virial, substructured morphology (NGC 1333f) akin to observed young star-forming regions. The latter is more dynamically active.

    The main results are:
    \begin{itemize}
        \item Captured asteroids are dynamically hotter than native ones (figure \ref{Fig:Heatmap_NGC1333f}). The same holds for planets (figure \ref{Fig:CaptPlanets_NGC1333f}).
        \item Clusters with more dynamical interactions produce larger numbers of interstellar objects and exchanged material, particularly around massive stars (figure \ref{Fig:AstFate_SpectralTypes_NGC1333f}).
        \item Bodies on wide orbits ($a>200$ au) are difficult to produce without external perturbations. External perturbations can seed long-term dynamical instabilities, allowing objects to scatter outwards, or can directly inject bodies onto wide orbits.
        \item Stellar encounters in cluster environments produce extreme trans-Neptunian object analogues, with the sub-virial, initially fractal configuration being more efficient than virialised Plummer models. In the former configuration, captured objects make up a substantial proportion of the extreme trans-Neptunian population, whereas in the latter, the population is predominantly native (figure \ref{Fig:Heatmap_NGC1333f}).
        \item The low Oort Cloud formation rate around Solar-like stars while still embedded in the cluster argues against the idea that the Oort Cloud has a predominantly captured asteroid population.
    \end{itemize}

    This work focuses on low-$N$ systems, which, while common \citep{2003ARA&A..41...57L}, have lower $N$ than that predicted for the Sun’s birth cluster \citep[$N\geq2000$, ][]{2009ApJ...696L..13P, 2010ARA&A..48...47A, 2023A&A...670A.105A}. Denser clusters tend to experience hyperbolic encounters rather than the parabolic ones characteristic of low-$N$ systems \citep{2006ApJ...642.1140O, 2010A&A...509A..63O, 2016ApJ...828...48V}. Although hyperbolic encounters are less disruptive than parabolic ones, they are more frequent, and their cumulative effect leads to higher ejection rates and asteroid exchange \citep{2010A&A...509A..63O, 2023MNRAS.523.4801W, 2024MNRAS.533.4485W}.
    
    \citet{2010A&A...509A..63O} found that cluster density has a weak influence on disc destruction as long as the stellar number density remained below $\rho_* \lesssim 2.1\times10^{4}$ pc$^{-3}$. After $5$ Myr of integration, the authors reported a native-disc survival fraction of $f_{\rm Nat}\approx0.85$ in this regime when probing down to $\rho_*\approx1.3\times10^{3}$ pc$^{-3}$, comparable to our value of $f_{\rm Nat}\approx0.82^{+0.01}_{-0.00}$ where $\rho_*\approx3\times10^{2}$ pc$^{-3}$. When the density increases to $\rho_*=4.2\times10^{4}$ pc$^{-3}$, the survival rate of the native population falls to $f_{\rm Nat}\approx0.4$. Similarly, in slightly longer integrations (10 Myr), \citet{2006A&A...454..811P} obtained $f_{\rm Nat}\approx0.75$ for initially virialised Plummer clusters with $\rho_*\approx5.3\times10^{3}$ pc$^{-3}$. The results of these two previous investigations combined with our own ($\rho_* \approx 250$ M$_\odot$ pc$^{-3}$) suggest that below some critical threshold, which itself lies between $10^{4}<\rho_*$ [pc$^{-3}$] $<2.1\times10^{4}$, encounter-driven disc erosion saturates and becomes largely insensitive to further density changes.
    
    Plummer models remain attractive due to their analytical simplicity and the expectation that substructure dissipates within a few crossing times \citep{2004A&A...413..929G, 2014MNRAS.438..620P}. However, both observations and theory show that stars form in clumpy environments \citep{1998MNRAS.297.1163B, 2009ApJ...696.2086S, 2010A&A...518L.102A, 2022MNRAS.515.2266A}. Our results show that even short-lived violent relaxation imprints lasting signatures on minor body architectures, implying that initial conditions must be considered in studies of planetary system and cluster evolution. Whether dense, virialised Plummer clusters produce similar effects to sparser, sub-virial fractal systems remains unclear, although comparisons with \citet{2024MNRAS.533.4485W} suggest possible similarity. Resolving this may help constrain the Solar System’s birth environment through minor body populations.

\section{Energy consumption}
    The \texttt{Nemesis} code exhibits variable CPU usage. Assuming $16$ cores are running and that, on average, a run takes $500$ hours, we obtain a final wall clock time of $16\times10\times500=80000$ hr. The cluster uses Intel Xeon Gold 5220R processors that use $150$ W. Given this, the total energy used is $E\approx 80000\times150\times\frac{1}{24}=500$ kWh. 
    
    In 2025, the Netherlands had a CO$_2$ equivalent of $0.388$ kg kWh$^{-1}$, of which $28\%$ of the energy produced was renewable\footnote{https://www.nowtricity.com/country/netherlands/}. Accounting for this green energy, the kWH to CO$_2$ equivalent emission is $0.27936$ kg kWh$^{-1}$. As such, the total equivalent CO$_2$ emission is $E_{\rm CO_2}\approx (500\, {\rm kWh})\times(0.27936 {\rm\, kg\, kWh^{-1}})\approx140$ kg. This roughly amounts to a round trip using a small diesel car between Leiden, the Netherlands and Arras, France ($\sim600$ km).
 
\begin{acknowledgements}
    The authors would like to thank the referee for their suggestions which helped improve the manuscript.
    
    Analysis was made using the open-source \texttt{Python} packages \texttt{NumPy} \citep{2020Natur.585..357H} and \texttt{Matplotlib} \citep{2007CSE.....9...90H}.
\end{acknowledgements}

\bibliographystyle{aa}
\bibliography{references.bib}

\begin{appendix}
    \section{Individual heatmaps}\label{Sec:AppA}  
\begin{figure*}[!htbp]
            \centering                
            \includegraphics[width=\columnwidth]{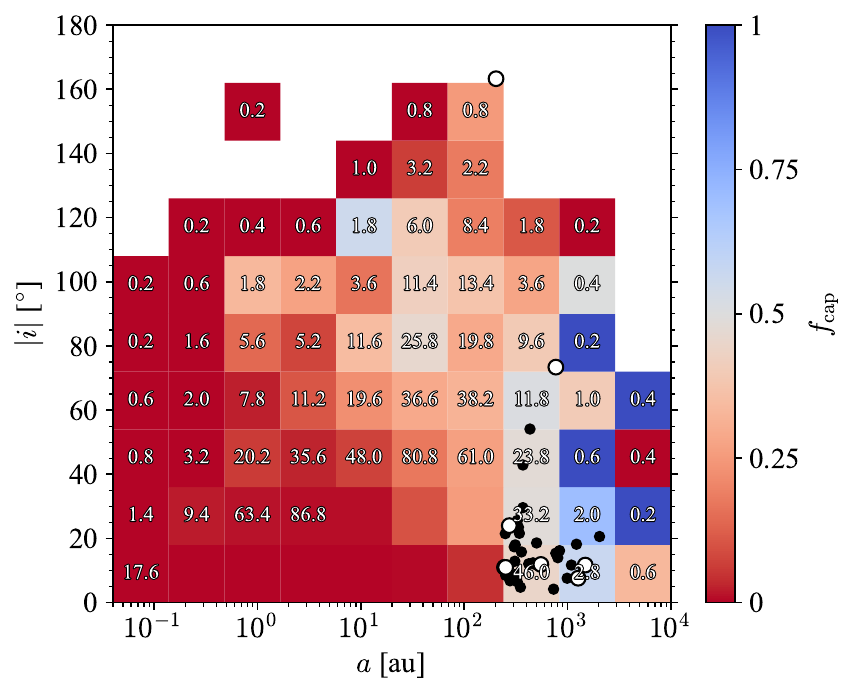}
            \includegraphics[width=\columnwidth]{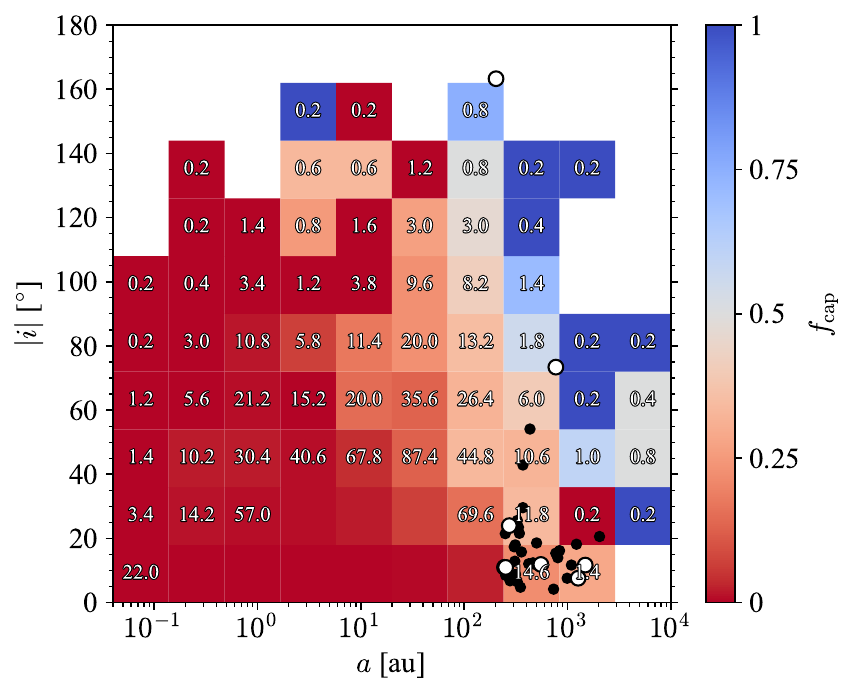}
            \includegraphics[width=\columnwidth]{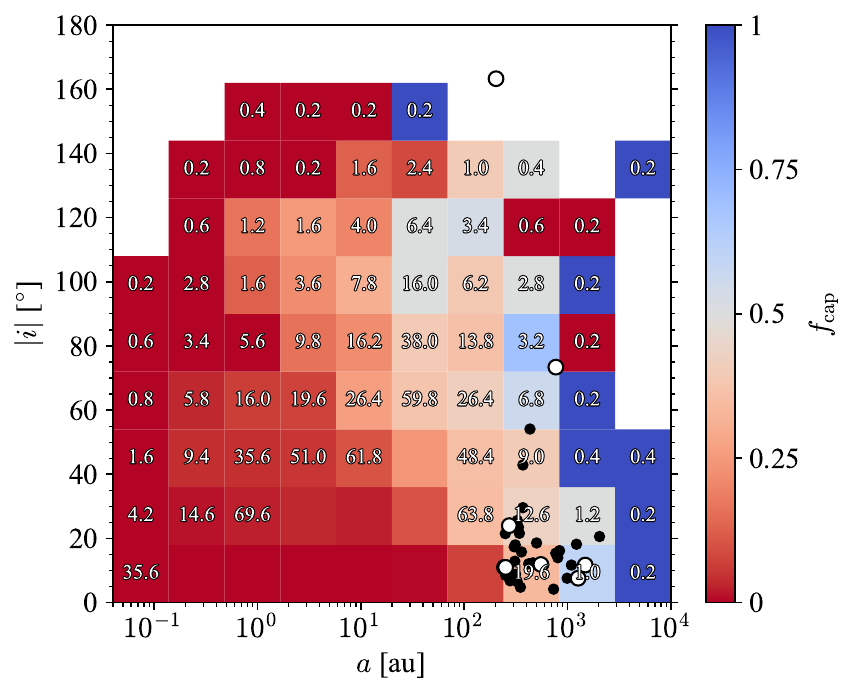}
            \includegraphics[width=\columnwidth]{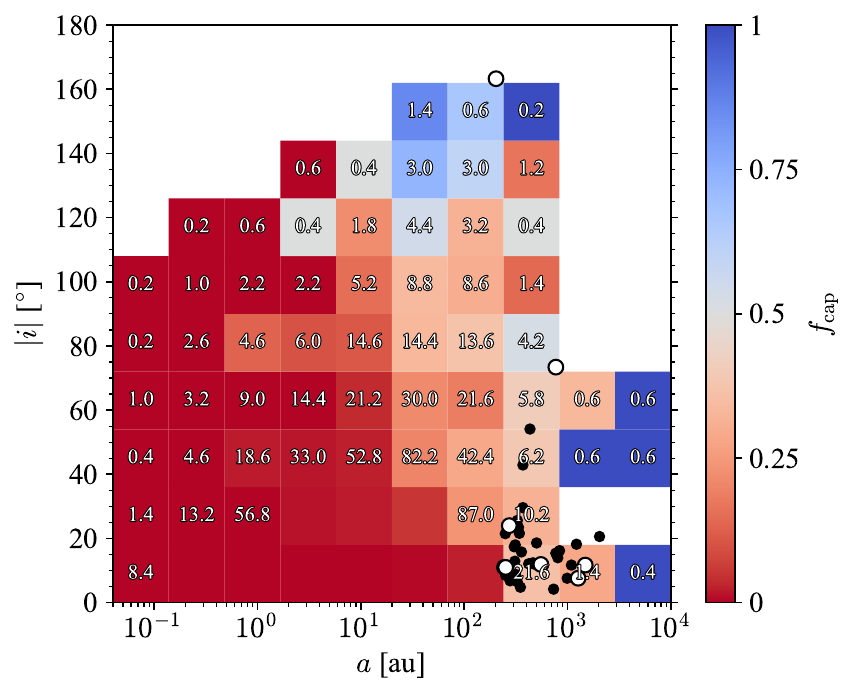}
            \includegraphics[width=\columnwidth]{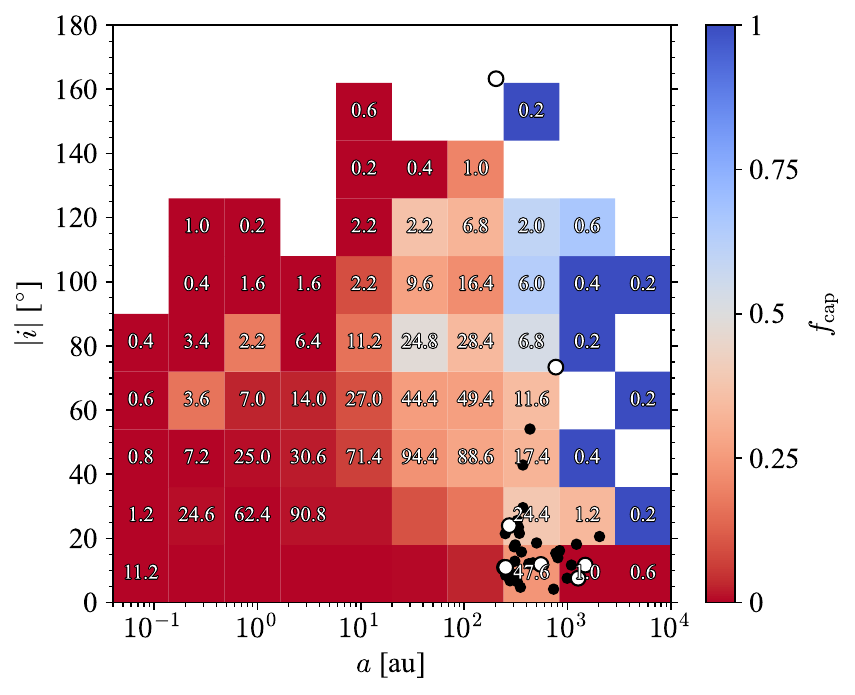}
            \caption{Bound asteroid's orbital semi-major axis versus eccentricity after $30$ Myr of integration. Coloured tiles represent the fraction captured, while numbers the population occupying some region. Each panel represents an individual run of NGC 1333f. Maximum cluster densities are: $\rho_{\rm core,\, max}=953$ M$_\odot$ pc$^{-3}$ (top left), $361$ M$_\odot$ pc$^{-3}$ (top right), $405$ M$_\odot$ pc$^{-3}$ (middle left), $84$ M$_\odot$ pc$^{-3}$ (middle right) and $600$ M$_\odot$ pc$^{-3}$ (bottom).}
            \label{Fig:HEAT1}
        \end{figure*}

\begin{figure*}[!htbp]
            \centering                
            \includegraphics[width=\columnwidth]{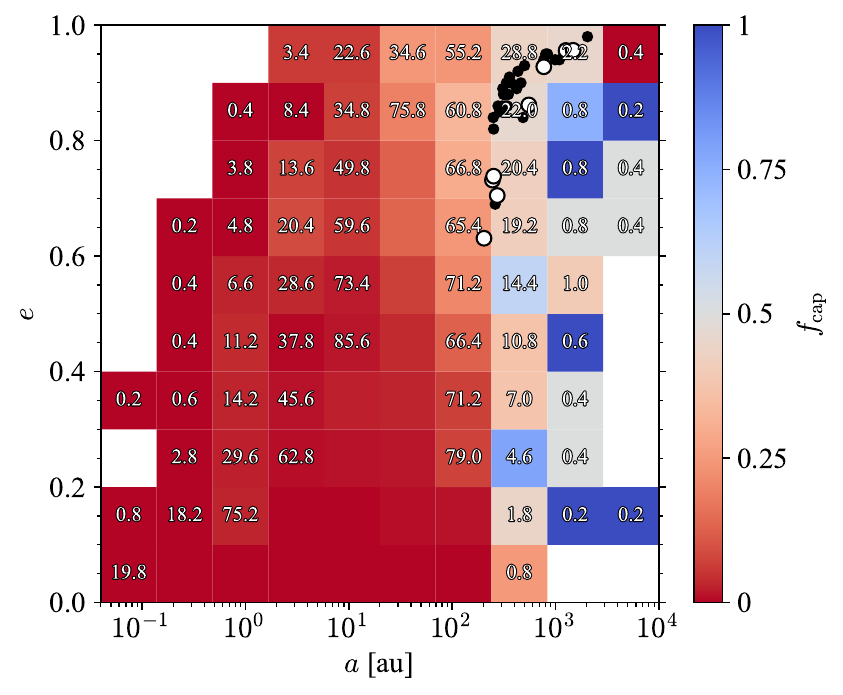}
            \includegraphics[width=\columnwidth]{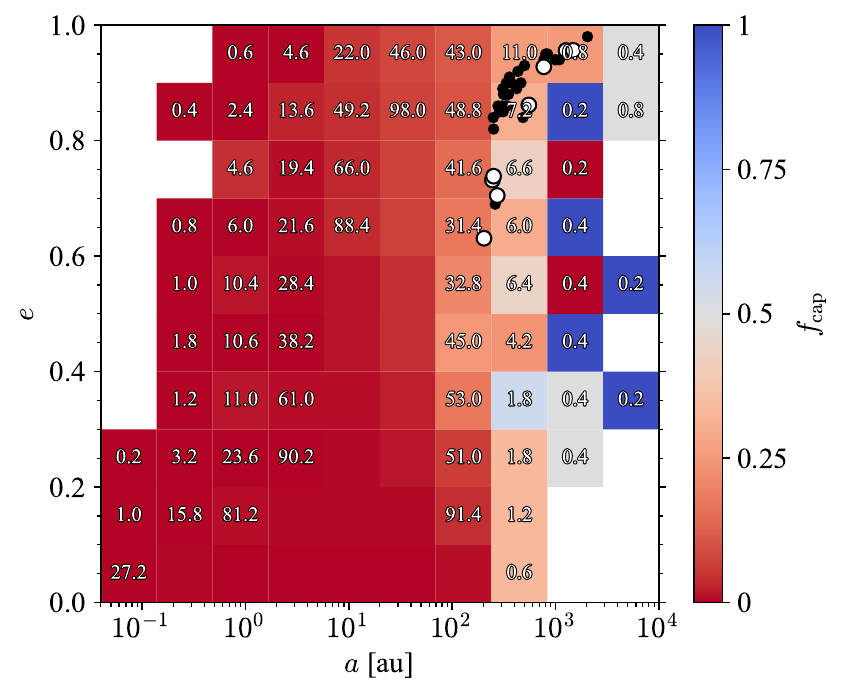}
            \includegraphics[width=\columnwidth]{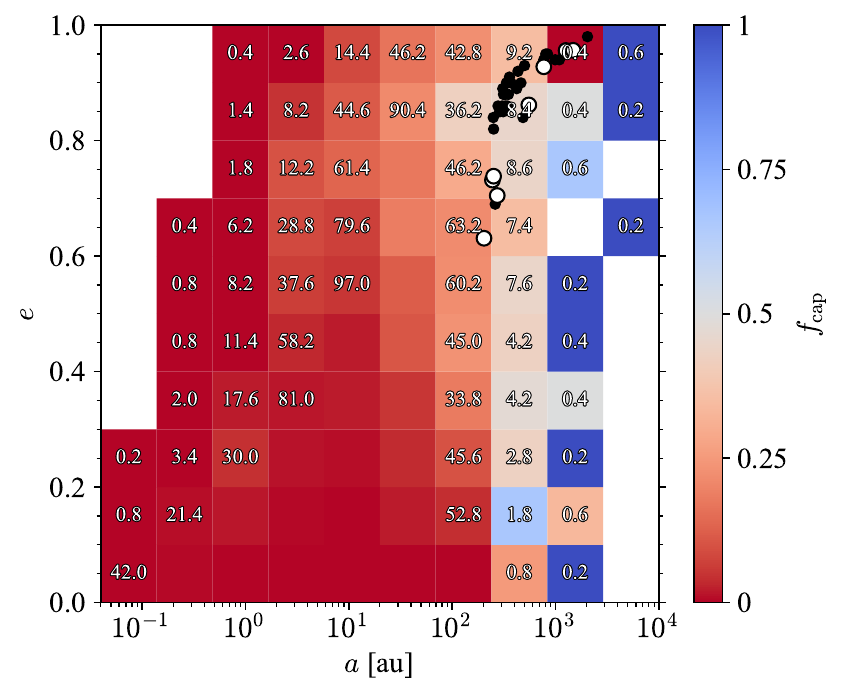}
            \includegraphics[width=\columnwidth]{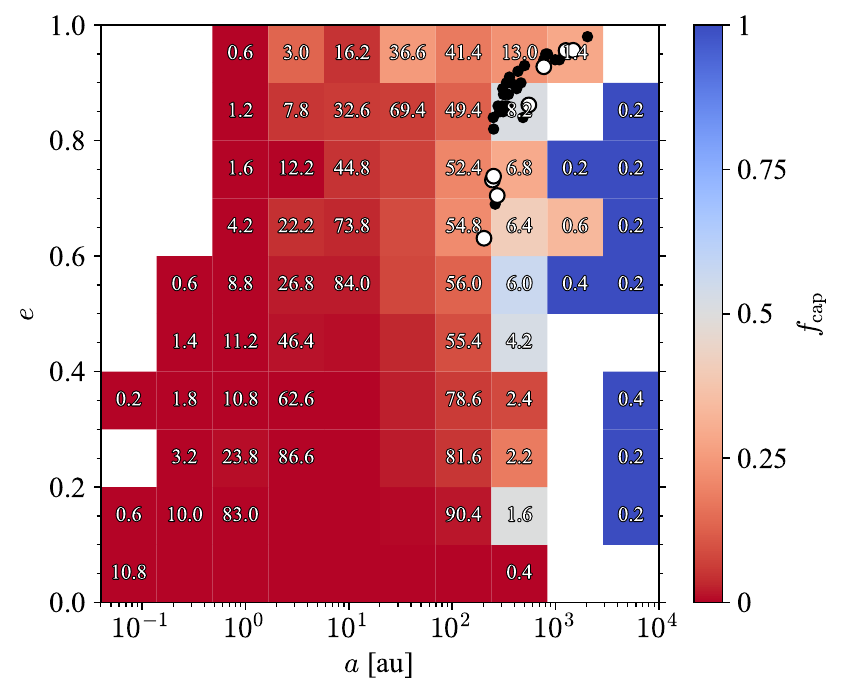}
            \includegraphics[width=\columnwidth]{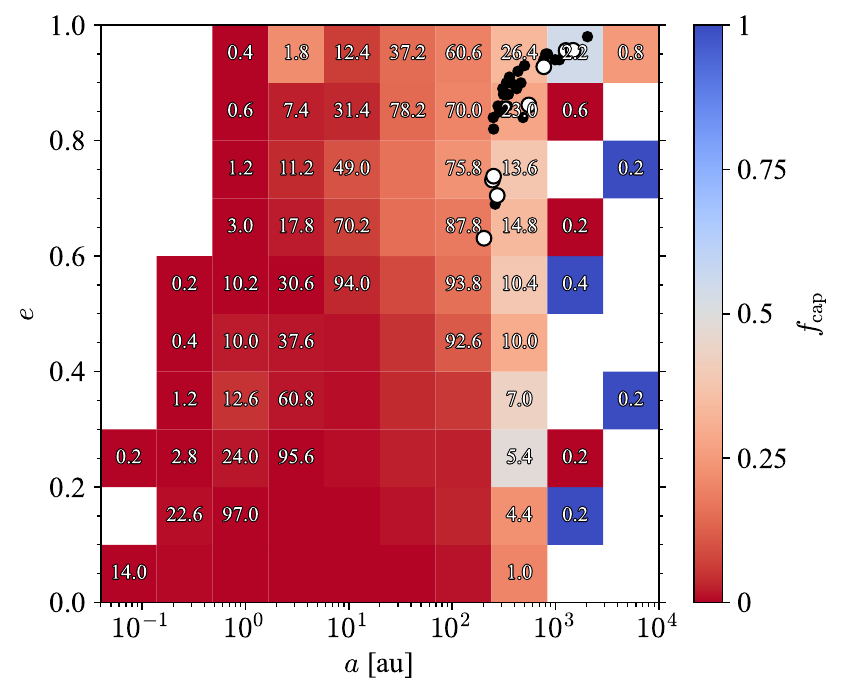}
            \caption{Bound asteroid's orbital semi-major axis versus inclination after $30$ Myr of integration. Coloured tiles represent the fraction captured, while numbers the population occupying some region. Each panel represents an individual run of NGC 1333f.}
            \label{Fig:HEAT2}
        \end{figure*}

\begin{figure*}[!htbp]
            \centering                
            \includegraphics[width=\columnwidth]{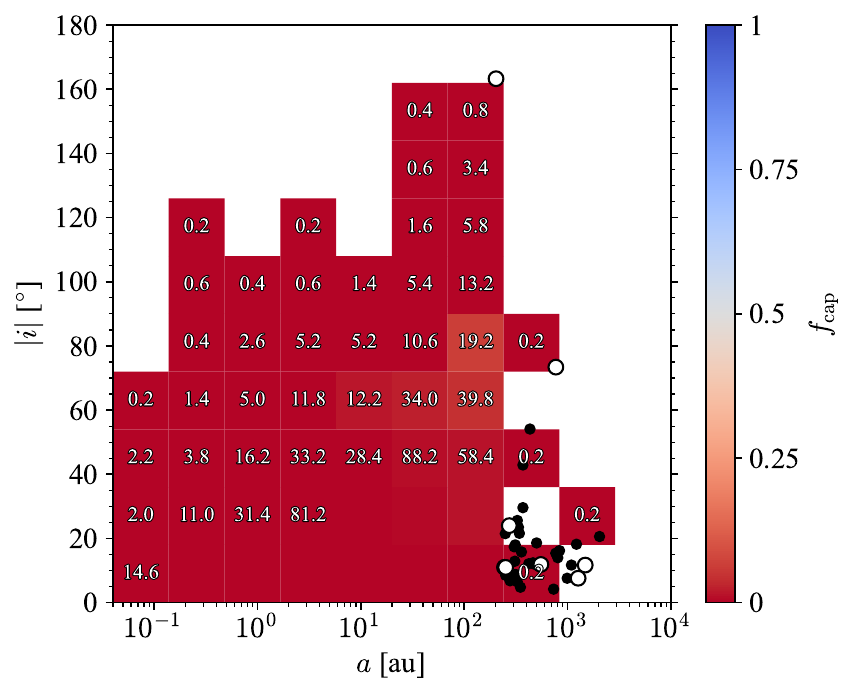}
            \includegraphics[width=\columnwidth]{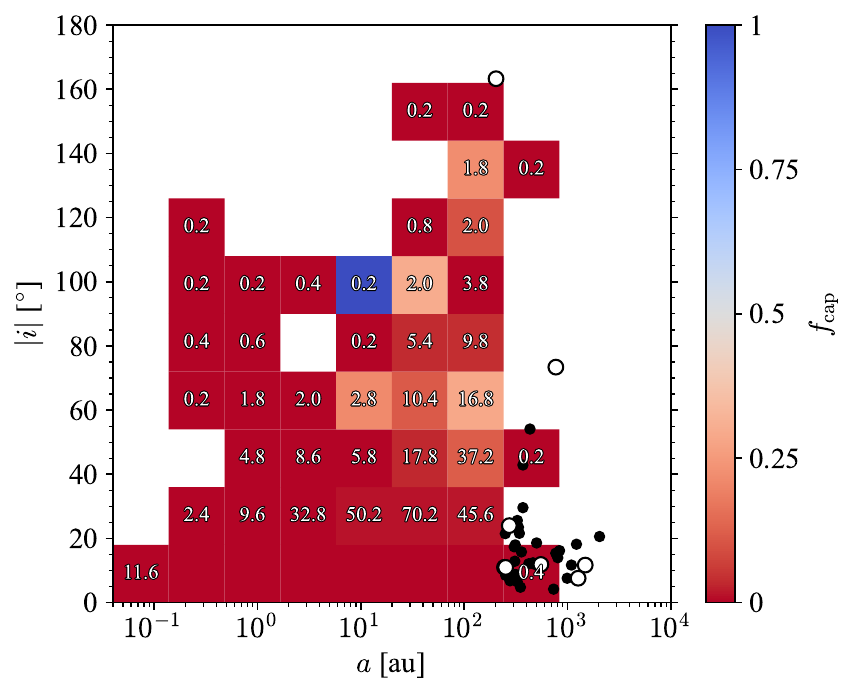}
            \includegraphics[width=\columnwidth]{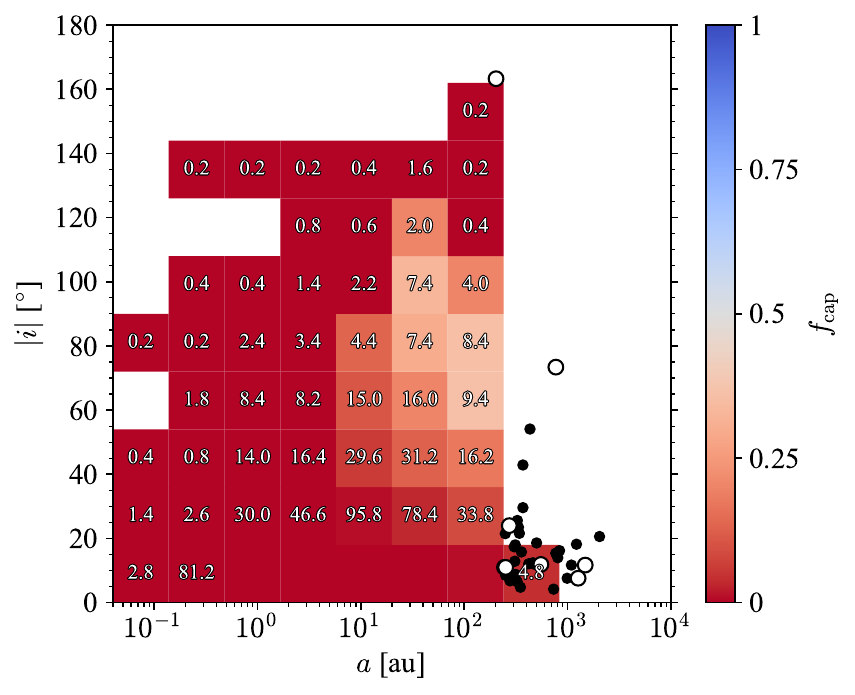}
            \includegraphics[width=\columnwidth]{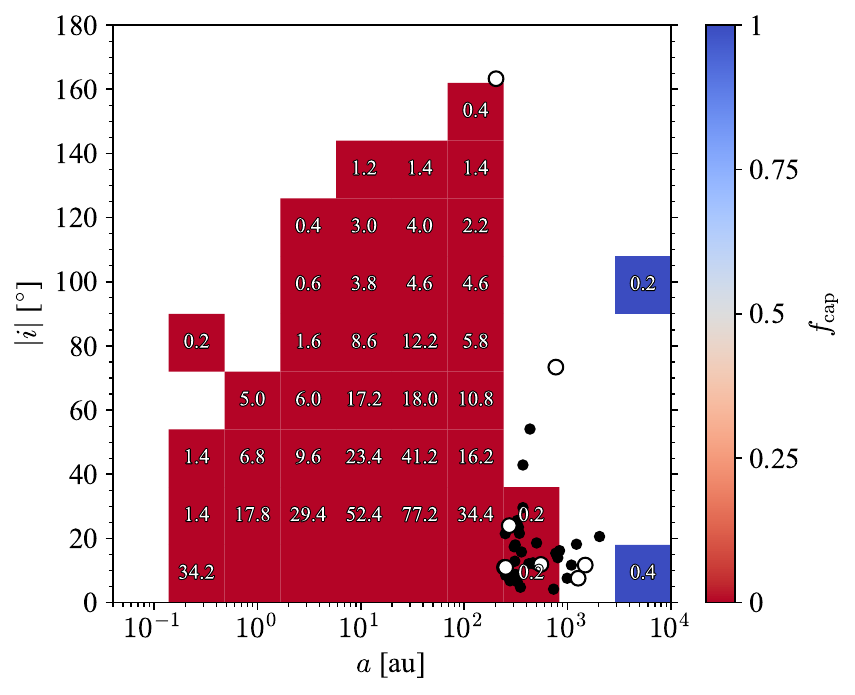}
            \includegraphics[width=\columnwidth]{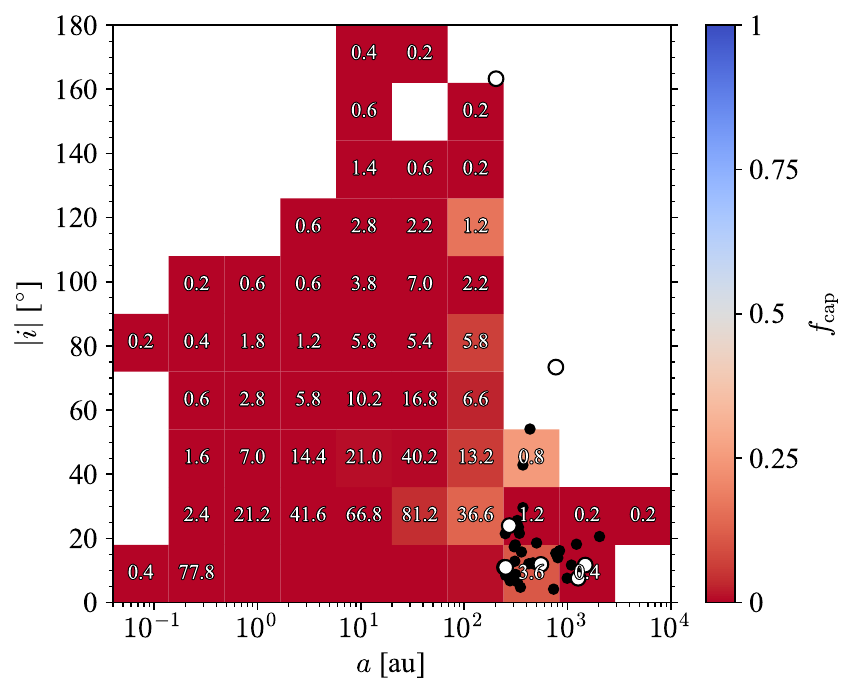}
            \caption{Bound asteroid's orbital semi-major axis versus eccentricity after $30$ Myr of integration. Coloured tiles represent the fraction captured, while numbers the population occupying some region. Each panel represents an individual run of NGC 1333p. Bound asteroid's orbital semi-major axis versus eccentricity after $30$ Myr of integration. Coloured tiles represent the fraction captured, while numbers the population occupying some region. Each panel represents an individual run of NGC 1333f. Maximum cluster densities are: $\rho_{\rm core,\, max}=1010$ M$_\odot$ pc$^{-3}$ (top left), $1030$ M$_\odot$ pc$^{-3}$ (top right), $1600$ M$_\odot$ pc$^{-3}$ (middle left), $810$ M$_\odot$ pc$^{-3}$ (middle right) and $2120$ M$_\odot$ pc$^{-3}$ (bottom).}
            \label{Fig:HEAT3}
        \end{figure*}

\begin{figure*}[!htbp]
            \centering                
            \includegraphics[width=\columnwidth]{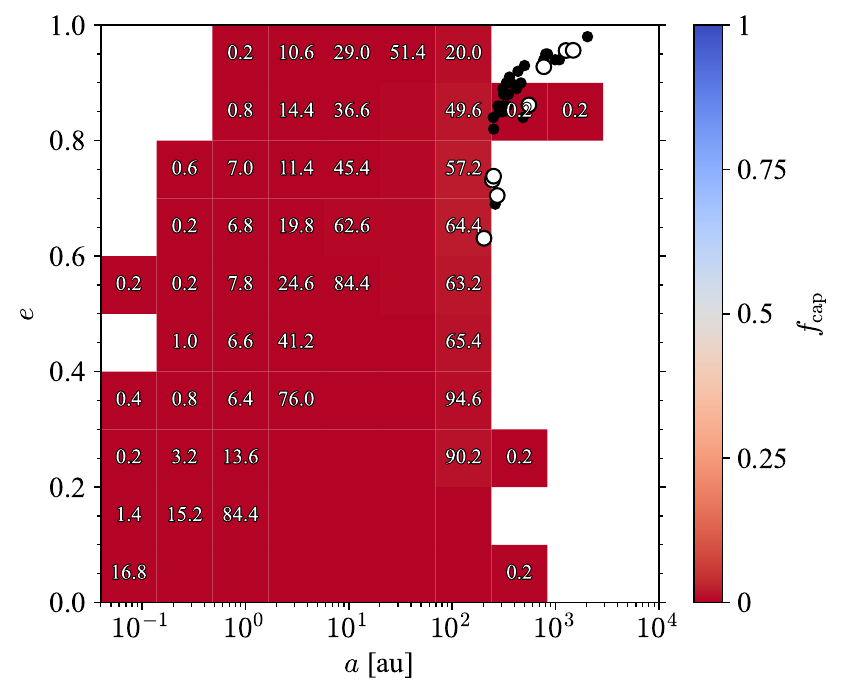}
            \includegraphics[width=\columnwidth]{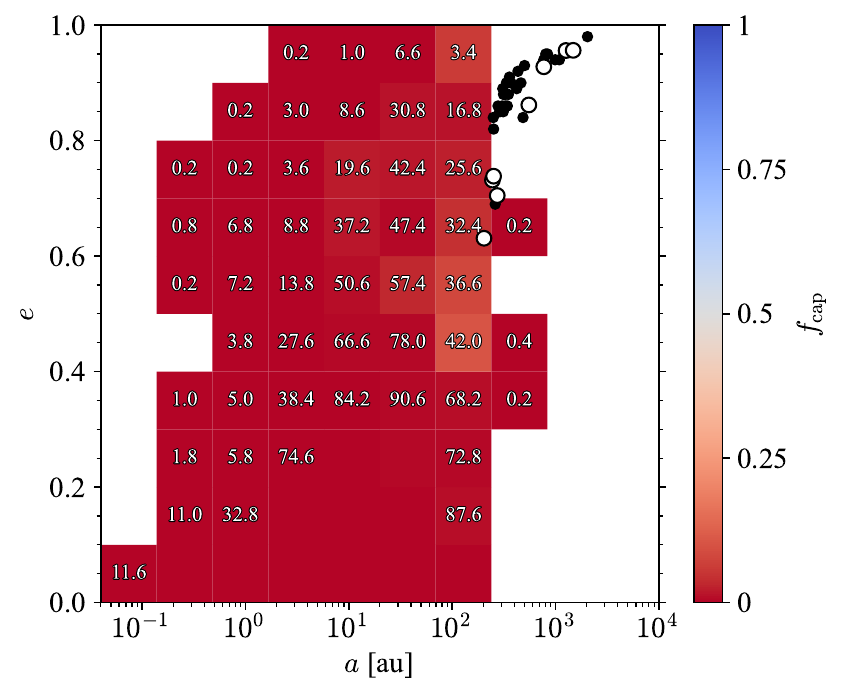}
            \includegraphics[width=\columnwidth]{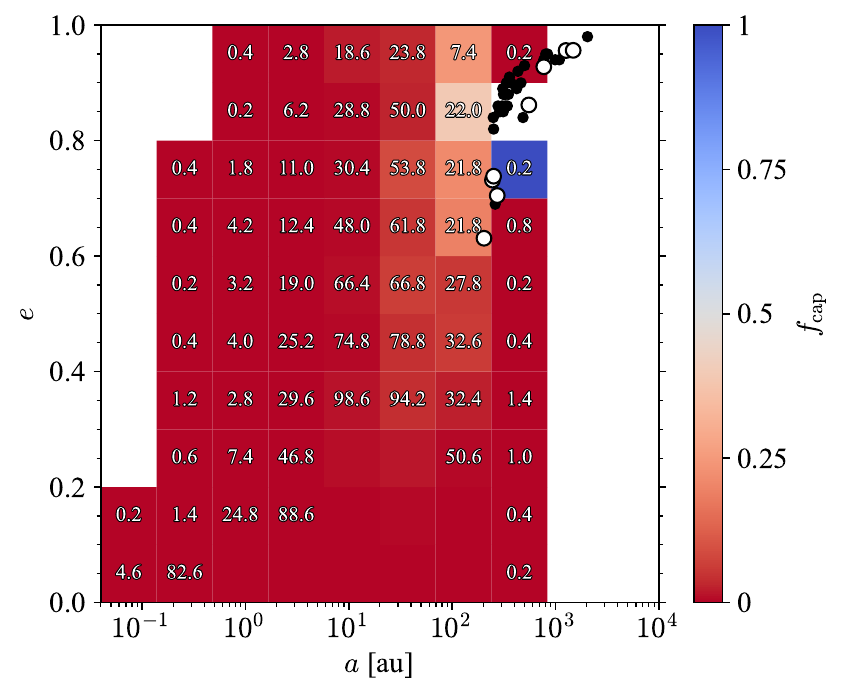}
            \includegraphics[width=\columnwidth]{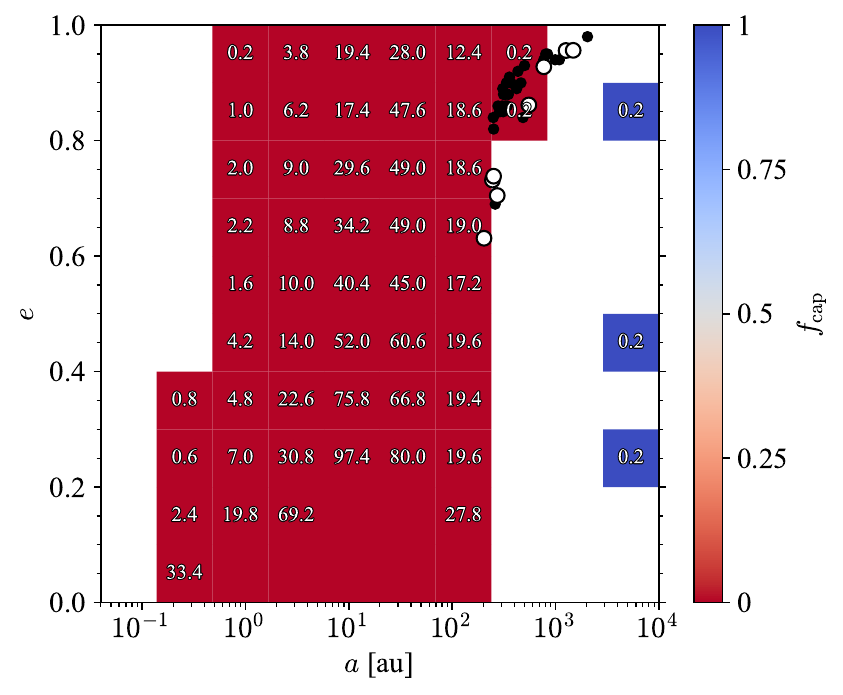}
            \includegraphics[width=\columnwidth]{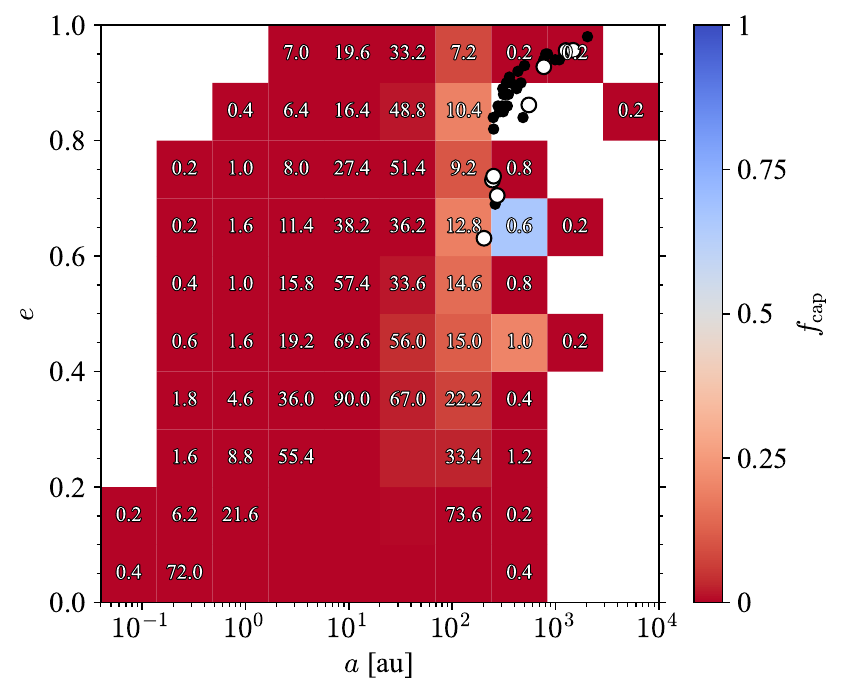}
            \caption{Bound asteroid's orbital semi-major axis versus inclination after $30$ Myr of integration. Coloured tiles represent the fraction captured, while numbers the population occupying some region. Each panel represents an individual run of NGC 1333p.}
            \label{Fig:HEAT4}
        \end{figure*}

        Figures \ref{Fig:HEAT1}, \ref{Fig:HEAT2},  \ref{Fig:HEAT3},  \ref{Fig:HEAT4} show where in $(a,i)$- and $(a,e)$-space bound asteroids lie in the individual runs. The former two figures are for NGC 1333f while the latter two, NGC 1333p.
        
        Within each model, the overall distributions is consistent across runs. In particular, all panels show a progressively dynamically hotter population at larger semi-major axes, with both inclination and eccentricity increasing with $a$. This reflects the greater susceptibility of wide orbits to perturbations in the cluster environment. In addition, the fraction of captured objects increases toward larger $a$, indicating that capture processes preferentially populate the outer regions of planetary systems.

        Stochasticity does arise, particularly at the extremities, indicating the sensitivity of the populations occupying this regions with the encounter history. Differences also appear at small semi-major axes and runs with higher peak core densities ($\rho_{\rm core,, max}$) show enhanced populations at low $a$. This is most clear when comparing the middle-right and bottom panels of Figures \ref{Fig:HEAT3} and \ref{Fig:HEAT4} though a similar trend also appears for NGC 1333f (top left panel versus middle right of figures \ref{Fig:HEAT1} and \ref{Fig:HEAT2}). This suggests that, in systems where the inner regions are otherwise depleted, external perturbations can act as an efficient mechanism for replenishing the small-$a$ population either by seeding internal instabilities or via direct injections.

        In all cases, while NGC 1333f has smaller maximum core density than NGC 1333p, its fractal-like distribution means encounters can be much more prominent and nearer one another (see e.g. figure \ref{Fig:TwoPoint}). As such, statistics on the minor bodies within the Solar System alone, including the Sednoids, is not enough to constrain the core density of the Sun's birth cluster as a degeneracy with virial ratio and initial distribution exists.
        

\clearpage
        
    \section{Statistical noise}
        Simulations here consider $500$ and $1000$ asteroids per debris disc for models NGC 1333f and NGC 1333p respectively. This is already several orders lower than the number of $>1$ km sized objects expected to be discovered over the course of the $10$ year LSST \citep[$\sim5.4\times10^{6}$,][]{2019ApJ...873..111I} which itself is limited by observational capabilities. With such small number statistics per disc considered here, results are susceptible to statistical noise, here we illustrate this with two examples.
        
\begin{figure}
            \centering                
            \includegraphics[width=\columnwidth]{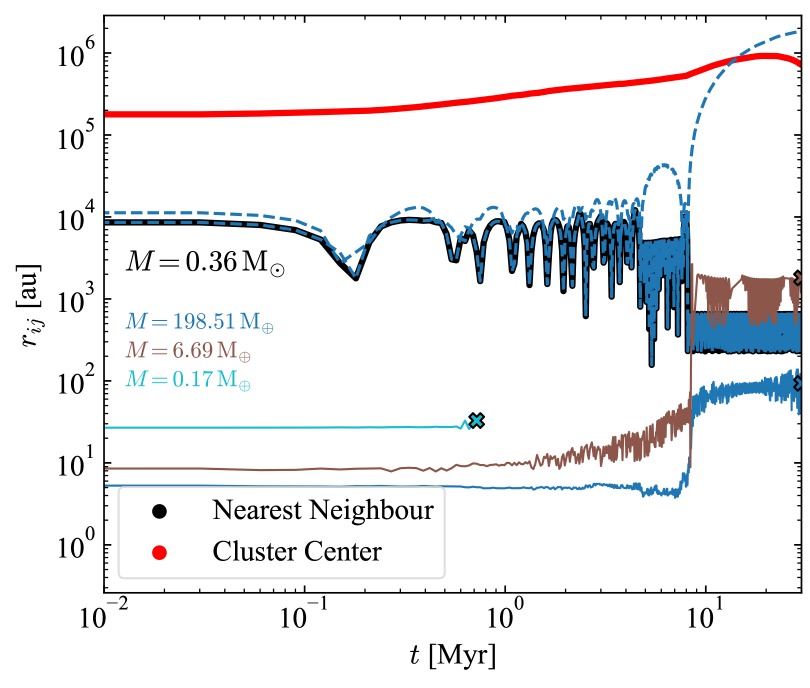}
            \includegraphics[width=\columnwidth]{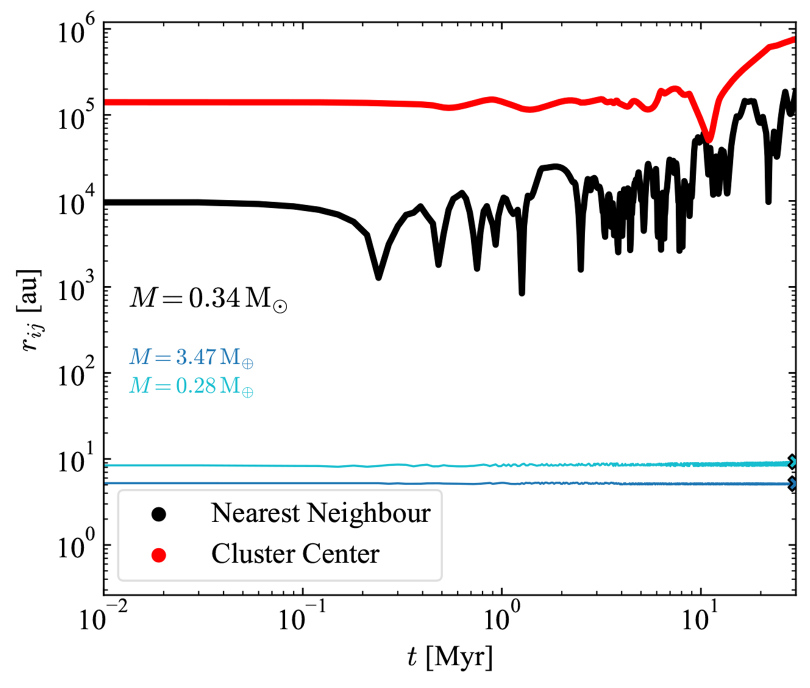}
            \caption{System history for select stars. Red solid lines show the distance between the star and the cluster centre, and in black the distance between the star to its nearest stellar neighbour. Coloured solid lines show the position of planets relative to the star. Coloured dashed lines show the position of an external star whose asteroids have been captured. The truncated line for the cyan-coloured planet in the top panel indicates the moment it has been ejected from the system.}
            \label{Fig:HistoryA}
        \end{figure}

        Figure \ref{Fig:HistoryA} shows the history of two planetary systems whose star have similar masses. The system on the top panel experiences a violent history, with a planet being ejected at $\sim0.8$ Myr, and a binary stellar system forming at $8$ Myr. While interactions with the external environment trigger migration within the planetary system, as shown with the gradual increase in $r_{ij}$ for the $6.69$ M$_\oplus$ planet, the instant the binary star has settled, the planetary orbits experience roughly an order-of-magnitude increase in their semi-major axis. The system on the bottom panel experiences a very different evolutionary history. Although it has $r_{ij}\approx10^{3}$ au approaches with other stars, these do not seem to influence the planetary orbits. The small oscillation in $r_{ij}$ imply that both planets also have low $e$. 

        \begin{figure}
            \centering                
            \includegraphics[width=.98\columnwidth]{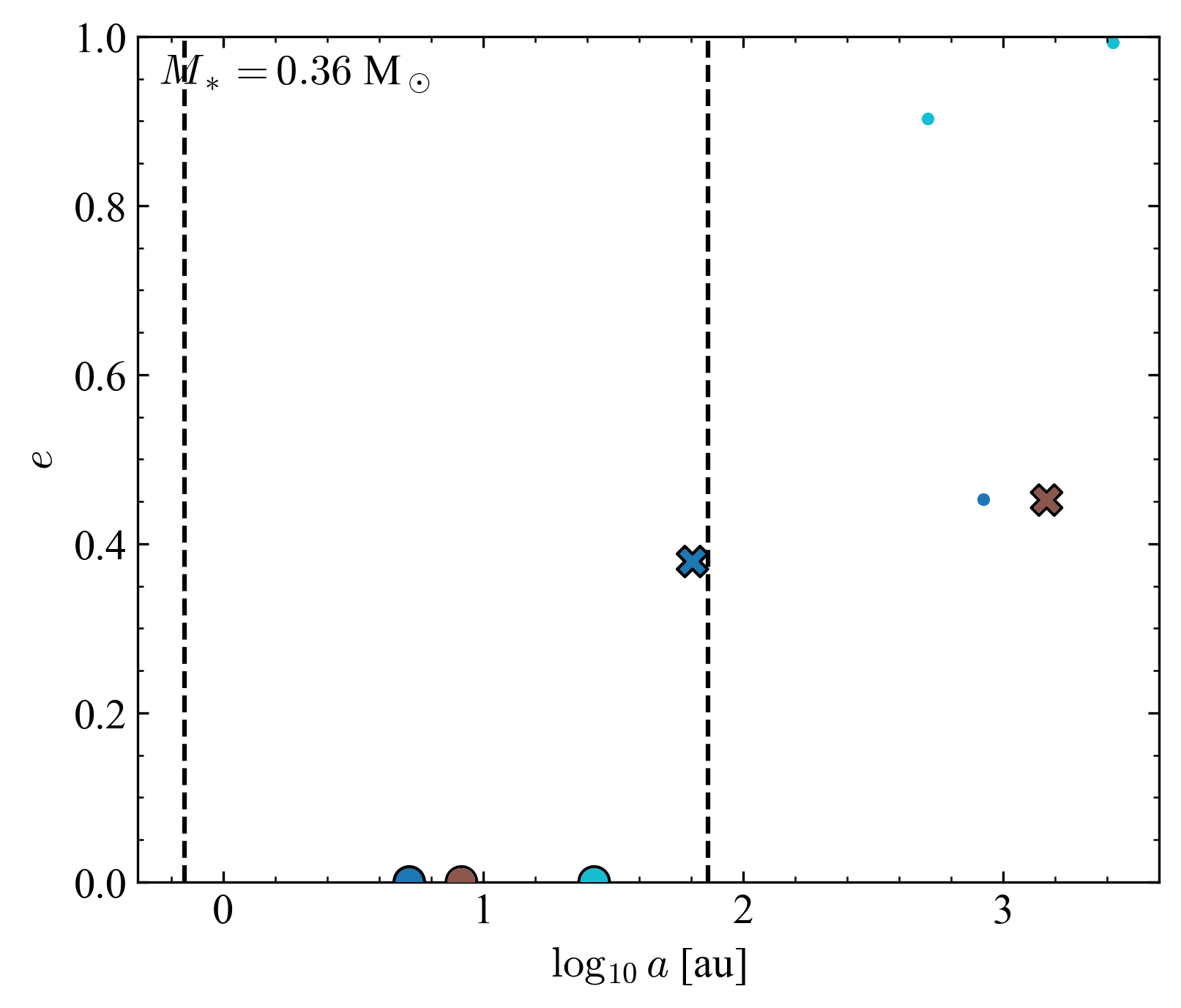}
            \includegraphics[width=.98\columnwidth]{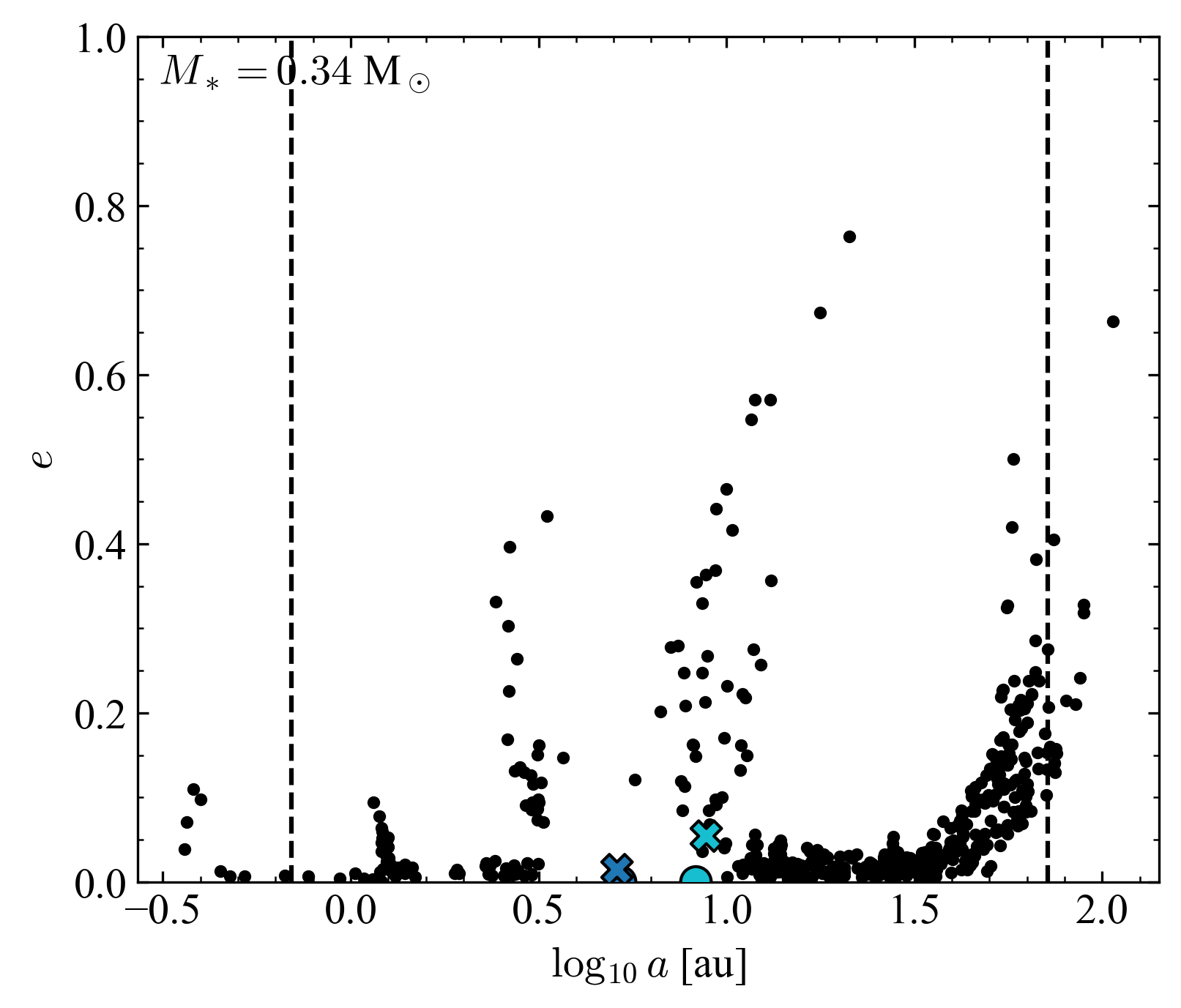}
            \vspace{0.6em}\caption{Semi-major axis and eccentricity of the remaining bound minor bodies. Small points represent asteroids, with black representing the native population, and each unique colour representing a different foreign origin. Large dots and crosses represent the initial and final planetary parameters, respectively.}
            \label{Fig:HistoryB}
        \end{figure}
        This difference in dynamical history between the two planetary systems imprints itself on their final asteroid population (figure \ref{Fig:HistoryB}). The dynamically active system ends with only three asteroids, with all having been captured. Meanwhile, the dynamically inert system has a more prominent debris disc but contains no captured asteroids. This system even exhibits the emergence of the Oort Cloud conveyor belt, with the presence of the high-$e$ asteroids near planets. 
        
        These examples highlight a limitation of small-number statistics. The dynamically active system in the top panel ends with $f_{\rm cap}=1.0$, whereas the bottom one has $f_{\rm cap}=0.0$. Overall ejections are more common than captures during an interaction (see also \citet{2018MNRAS.473.5432H}) making systems with an eventful past having a near depleted debris disc. The low number of remaining asteroids results in any capture event inflating $f_{\rm cap}$. Yet, in the cumulative distribution function shown in Fig. \ref{Fig:ExoticPop_NGC1333f}, each star contributes equally, regardless of asteroid count. This equal weighting can skew the inferred fraction of stars with $f_{\rm cap}$. 
        
        One alternative is to compute the capture fraction over all individual asteroids rather than averaging per star, which reduces the influence of extreme outliers. However, this approach suppresses the system-to-system variation illustrated here and tends to bias the result toward systems who had little-to-no stellar encounters since they tend to better preserve their discs. Our choice of considering system-to-system variation rather than bulk population statistics is motived since this helps us better understand which stellar populations are more likely to contain captured objects and also provides a rough idea on their expected proportions across spectral types.

        Lastly, in dynamical systems containing $N>2$ bodies, the history of the system is extremely sensitive to initial conditions \citep{1891BuAsI...8...12P, 1892mnmc.book.....P, 1986LNP...267..212D}. This sensitivity to initial conditions is made apparent here with similar systems having completely different outcomes. While global averages can provide insights, outliers exist, making it hard to give predictions on a specific system without thorough knowledge of its dynamical past. This highlights the complication in linking results back to the Solar System, whose birth environment is only roughly constrained, and whose dynamical history is even less so. The validity of results here with the Solar System in particular is further complicated by the modelled cluster likely being less massive and dense than that with which the Sun originated from \citep{2009ApJ...696L..13P, 2010ARA&A..48...47A}.
        
\end{appendix}

\end{document}